\documentclass[twocolumn]{aastex62}
\usepackage{color}

\begin{document}

\title{
    First observation of MeV gamma-ray universe with bijective imaging spectroscopy \\
    using the Electron-Tracking Compton Telescope aboard SMILE-2+
}

\correspondingauthor{Atsushi Takada}
\email{takada@cr.scphys.kyoto-u.ac.jp}

\author{Atsushi Takada}
\author{Taito Takemura}
\author{Kei Yoshikawa}
\affil{
	Graduate School of Science, Kyoto University \\
	Kitashirakawa Oiwakecho, Sakyo, Kyoto, Kyoto, 606-8502, Japan
}

\author{Yoshitaka Mizumura}
\affil{
	Institute of Space and Astronautical Science, Japan Aerospace Exploration Agency \\
	Yoshinodai 3-1-1, Chuou, Sagamihara, Kanagawa, 252-5210, Japan
}

\author{Tomonori Ikeda}
\author{Yuta Nakamura}
\author{Ken Onozaka}
\affil{
	Graduate School of Science, Kyoto University \\
	Kitashirakawa Oiwakecho, Sakyo, Kyoto, Kyoto, 606-8502, Japan
}

\author{Mitsuru Abe}
\affil{
	Graduate School of Science, Kyoto University \\
	Kitashirakawa Oiwakecho, Sakyo, Kyoto, Kyoto, 606-8502, Japan
}

\author{Kenji Hamaguchi}
\affil{
	Department of Physics, University of Maryland, Baltimore County \\
	1000 Hilltop Circle, Baltimore, MD 21250, USA
}

\author{Hidetoshi Kubo}
\affil{
	Graduate School of Science, Kyoto University \\
	Kitashirakawa Oiwakecho, Sakyo, Kyoto, Kyoto, 606-8502, Japan
}

\author{Shunsuke Kurosawa}
\affil{
	Institute of Materials Research, Tohoku University\\
	Katahira 2-1-1, Aoba, Sendai, Miyagi, 980-8577, Japan
}

\author{Kentaro Miuchi}
\affil{
	Graduate School of Science, Kobe University \\
	1-1 Rokkoudai-cho, Nada-ku, Kobe, Hyogo, 657-8501, Japan
}

\author{Kaname Saito}
\affil{
	Graduate School of Science, Kyoto University \\
	Kitashirakawa Oiwakecho, Sakyo, Kyoto, Kyoto, 606-8502, Japan
}

\author{Tatsuya Sawano}
\affil{
	Graduate School of Natural Science and Technology, Kanazawa University \\
	Kakuma, Kanazawa, Ishikawa, 920-1192, Japan
}

\author{Toru Tanimori}
\affil{
	Graduate School of Science, Kyoto University \\
	Kitashirakawa Oiwakecho, Sakyo, Kyoto, Kyoto, 606-8502, Japan
}

\begin{abstract}
MeV gamma-rays provide a unique window for the direct measurement of line emissions from radioisotopes, 
but observations have made little significant progress after COMPTEL/{\it CGRO}. 
To observe celestial objects in this band, we are developing an electron-tracking Compton camera (ETCC), 
which realizes both bijective imaging spectroscopy
and efficient background reduction gleaned from the recoil electron track information.
The energy spectrum of the observation target can then be obtained
by a simple ON-OFF method using a correctly defined point spread function on the celestial sphere.
The performance of celestial object observations was validated on the second balloon SMILE-2+ installed with an ETCC having a gaseous electron tracker 
with a volume of 30$\times$30$\times$30~cm$^3$.
Gamma-rays from the Crab nebula were detected
with a significance of 4.0$\sigma$ in the energy range 0.15--2.1~MeV
with a live time of 5.1 h, as expected before launching.
Additionally, the light curve clarified an enhancement of gamma-ray events generated in the Galactic center region, 
indicating that a significant proportion of the final remaining events are cosmic gamma rays.
Independently, 
the observed intensity and time variation were consistent with the pre-launch estimates except in the Galactic center region.
The estimates were based on the total background of extragalactic diffuse, atmospheric, and instrumental gamma-rays after accounting for
the variations in the atmospheric depth and rigidity during the level flight. 
The Crab results and light curve strongly support our
understanding of both the detection sensitivity and the background in real observations.
This work promises significant advances in MeV gamma-ray astronomy.
\end{abstract}

\keywords{Gamma-ray astronomy --- Gamma-ray telescopes --- High altitude balloons}

\section{Introduction}
Various radiation process in the universe can be observed in the low-energy gamma-ray band (0.1--100~MeV).
Examples are line emissions from the radioisotopes produced by nucleosynthesis 
in supernovae or neutron star mergers \citep{1988Natur.331..416M, 1995ExA.....6...85V},
the electron--positron annihilation line in the Galactic center region (GCR) \citep{2011RvMP...83.1001P}, 
synchrotron emissions and inverse Compton scattering with particle acceleration 
in active galactic nuclei or gamma-ray bursts (GRBs) \citep{1995PASP..107..803U, 1999ApJ...524...82B},
pion-decay radiation in the strong gravity fields 
around black holes \citep{1996A&AS..120C.149M, 1997ApJ...486..268M},
and de-excitation lines from nuclei excited by interactions 
between cosmic-rays and the interstellar medium \citep{2000ApJ...544..320B, 2000ApJ...537..763S}.
Population III stars are expected to be detected 
as long duration GRBs \citep{2010ApJ...715..967M, 2011ApJ...731..127T},
because the universe is very transparent in this energy band.
Thermal emissions in the extragalactic diffuse emission might confirm the existance of
primordial black holes (PBHs) with masses of $10^{16-17}$~g,
which emit thermal emission in the MeV band \citep{2010PhRvD..81j4019C}.

Although celestial MeV gamma-rays have been observed scince the dawn of high-energy astrophysics,
when GRBs were discovered by {\it Vela} 
and extragalactic diffuse emission was detected by {\it Apollo 15} \citep{1973ApJ...181..737T},
observations in this band stagnate.
COMPTEL aboard {\it CGRO} \citep{1993ApJS...86..657S} discovered 
only $\sim$30 steady gamma-ray sources in the 0.75--30~MeV band \citep{2000A&AS..143..145S},
and SPI loaded on {\it INTEGRAL} discovered only four steady celestial objects 
at energies above 0.6 MeV \citep{2008ApJ...679.1315B}.
The expected signatures of supernova explosions are
line gamma-rays emitted from fresh isotopes, but
line gamma-rays of $^{56}$Ni/$^{56}$Co have been detected 
only from SN1987A \citep{1988Natur.331..416M} and SN2014J \citep{2014Sci...345.1162D, 2015ApJ...812...62C}.
Various gamma-ray telescopes are being developed and some balloon experiments have been performed,
but none of the present developments have surpassed the sensitivity of COMPTEL
\citep{2008NIMPA.593..414A, 2011ApJ...738....8B, 2011PhDT........75K}.
The NCT (predecessor of COSI) produced some observational results 
using a wide-view Ge Compton Camera loaded on a few balloon experiments.
In 2009, COSI detected Crab at the 4$\sigma$ \citep{2011ApJ...738....8B} significance level.
In 2016, they detected the annihilation line from GCR
at the 7$\sigma$ significance level \citep{2020ApJ...895...44K, 2020ApJ...897...45S}.

MeV observations have been bottlenecked by the huge background and difficulty of imaging.
Unlike visible light or X-rays, MeV gamma-ray wavelengths are is too short to focus 
by a mirror or lens.
Obtaining their total energy is also difficult 
because whereas most incident photons deposit only part of their energy via Compton scattering, 
Compton scattering domainates the interactions between MeV gamma-rays and materials.
In addition, observations are obstructed by huge amounts of background photons
produced by hadronic interactions between cosmic rays 
and the material surrounding the detector \citep{2001A&A...368..347W}.
The SPI spectrometer and other detectors based on coded aperture imaging 
infer the intensity map of incident gamma-rays from the pattern of the shadow image. 
Such detectors need many photons to obtain the directions of celestial objects.
Coded aperture imaging telescopes are usually equipped with an active veto counter,
but heavy anti-coincidence counters scarcely improve the signal to noise ratio 
because the gamma rays produced by cosmic-rays are delayed\citep{2018A&A...611A..12D}.
Conventional Compton cameras such as COMPTEL can slightly suppress the contaminating background photons
by partially restricting the incident direction of each photon.
However, this type of camera measures only one of two angles representing the direction of incident gamma-rays
due to the lack of recoil direction.
The signal-to-noise ratio depends on the volume of the cone-shape response in the Compton data space~\citep{1993ApJS...86..657S},
in which the information of recoil directions degenerates.
Therefore, the observations are obstructed by large contamination of gamma-rays 
within the area spanned by a radius defined by the average of the detectable scattering-angle.
The detection sensitivity of a non-bijection telescope 
based on coded aperture imaging or conventional Compton imaging is 
inherently restricted by the principle confusion limit caused 
by the overlapped responses of the surrounding sources.
In fact, \citet{2004NewAR..48..193S} argued 
that above all abilities, next-generation MeV gamma-ray telescopes must distinguish and reject the background
based on sharp point spread function (PSF) on the celestial sphere
and additional event-selection parameters.
In early 2000, 
the MEGA group measured the tracks of recoil electrons with enegies higher than 2 MeV
using a Si-tracker and CsI calorimeters.
They reported worse angular resolution measures (ARMs) of tracked events than untracked events 
because detecting a few sampling points of the electron track requires a high recoil energy\citep{2003NSSMIC.2003.1694Z}.
Thus, a tracking detector must finely track electrons, even those with very low energies ($\sim$10 keV) 
to obtain a sufficient PSF for MeV gamma-ray astronomy.

As the next MeV gamma-ray telescope for deep sky surveying,
we are developing an electron-tracking Compton camera (ETCC),
which applies a gaseous electron tracker as the Compton-scattering target 
and pixel scintillator arrays as the absorbers \citep{2004NewAR..48..263T}.
The ETCC and conventional Compton camera differ in their tracking of Compton-recoil electrons. 
The ETCC obtaines the momentum of an incident gamma-ray by simply summing
the momenta of the scattered gamma-ray and the recoil electron event-by-event.
In this way, it completely reconstructs the Compton scattering process.
Thus, (like other wavelength telescopes)
an ETCC is a bijection telescope, which obtains the incident direction as the zenith and azimuthal angles 
and forms a proper PSF on the celestial sphere. 
As the proper PSF defines the minimum size that conserves the intensity of gamma rays, 
the spectroscopic information at many points beyond the PSF can be independently obtained from one image.
Therefore, the ETCC can realize imaging spectroscopy in MeV gamma-ray astronomy.
The proper PSF on the celestial sphere enables  
the determination of the energy spectrum of the observation target by a simple ON-OFF method \citep{2017NatSR...741511T}.
Moreover, from the recoil-electron track,
the background can be rejected using two powerful tools \citep{2015ApJ...810...28T}:
particle identification based on the energy deposition rate $dE/dx$ in the gaseous electron tracker,
and a Compton-scattering kinematic test based on the angle 
between the directions of scattered gamma-ray and recoil electron.
These background rejection tools enables ETCC observations without a heavy veto counter;
consequently, the ETCC has a large field of view (FoV).
These unique abilities of ETCC provide a real imaging spectroscopy to MeV gamma-ray observations.
In future observations with ETCCs loaded on a satellite \citep{2019BAAS...51g.145H},
we are planning balloon experiments,
named Sub-MeV/MeV gamma-ray Imaging Loaded-on-balloon Experiments (SMILE).
As the first step, we launched a small ETCC having an electron tracker 
with a sensitive volume of 10$\times$10$\times$15~cm$^3$ in 2006.
This launch was intended as a background study at high altitudes (SMILE-I)
and confirmation of the background rejection power of the ETCC \citep{2011ApJ...733...13T}. 
SMILE-I successfully detected diffuse cosmic and atmospheric gamma-rays 
and performed powerful background rejection based on particle identification. 
As the second step, SMILE was tested on imaging-spectroscopy observations of bright celestial objects.
To this end, we set the Crab nebula and GCR as the observation targets,
and constructed a middle-size ETCC a sensitive volume of 30$\times$30$\times$30~cm$^3$. 
Assuming a background of extragalactic diffuse and atmospheric gamma rays, 
the Crab nebula should be detected at the 3--5$\sigma$ significance level.
The second balloon SMILE-2+ was launched at Alice Springs, Australia, on April 7 of 2018.

Herein, we assess the gamma-ray detection abilities of the ETCC from ground calibrations,
details of the SMILE-2+ flight, and the observation results of the Crab nebula.
We additionally discuss the detection sensitivities of next ETCC observations
by comparing the realized detection sensitivity with that estimated from ground calibrations.

\section{Instruments}
\subsection{SMILE-2+ ETCC and control system}
\begin{figure}
	\plotone{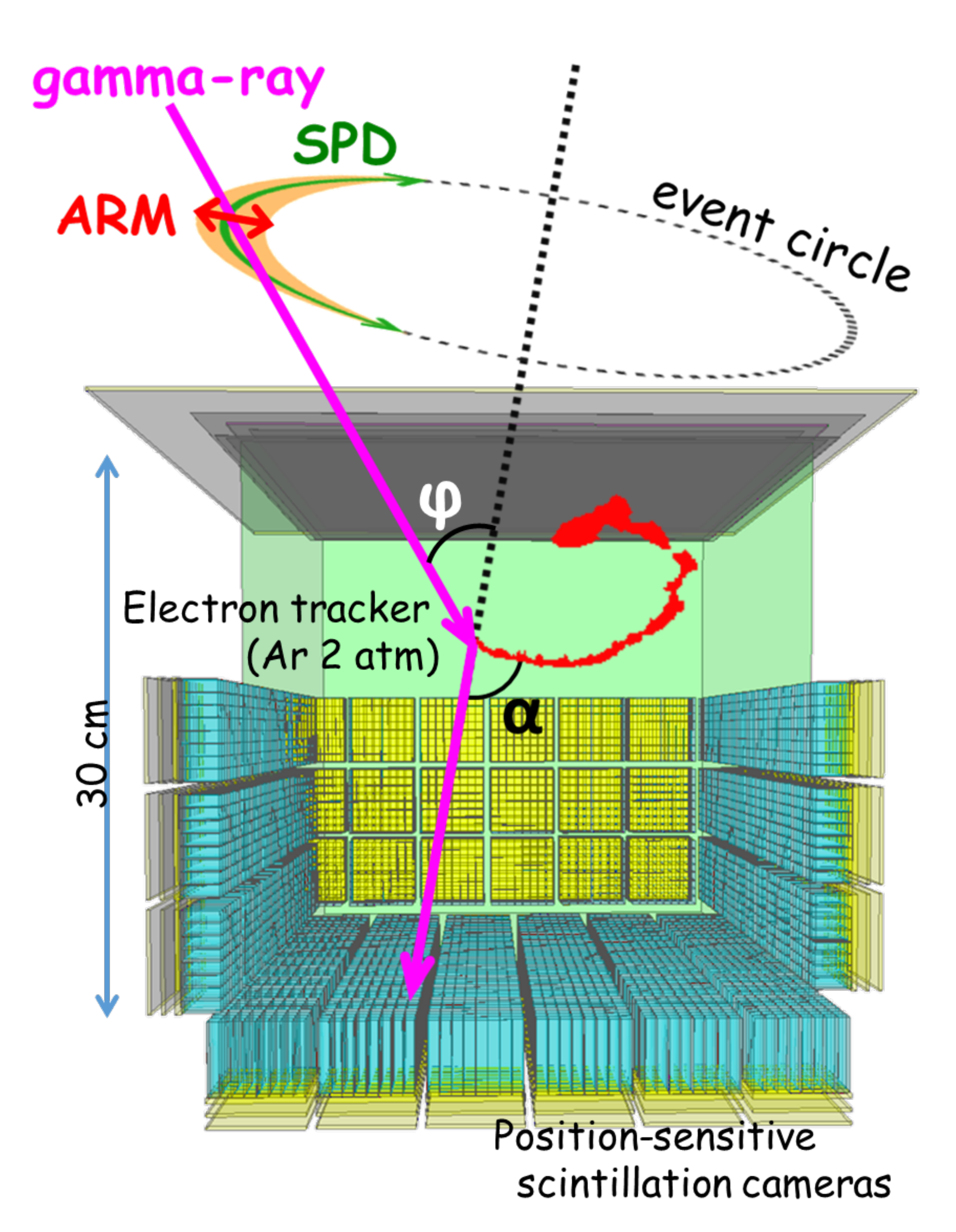}
	\caption{
		Schematic of the electron-tracking Compton camera (ETCC).
		The ETCC comprises a gaseous electron tracker as the Compton-scattering target
		and position-sensitive scintillation cameras that absorb the scattered gamma-rays.
	}
	\label{fig:etcc}
\end{figure}

At middle latitudes in the southern hemisphere,
the large zenith angle ($>$45 degrees) reduces
the flux of the Crab nebula by one half from that at low zenith angles;
consequently, the air mass is twice that at the zenith.
Thus, Crab nebula is difficult to detect even after several hours of balloon observations.
To detect the Crab nebula at the 3--5$\sigma$ significance level within during a few hours 
in the 0.2--2~MeV energy band at 40~km altitude in the southern hemisphere,
the ETCC requires a minimum effective area of $\sim$1~cm$^2$ (0.3~MeV). 
The required PSF is $\sim$30~degrees for 0.6~MeV detection at the half power radius (HPR),
and the instrumental background must be suppressed 
to below the background of the diffuse cosmic and atmospheric gamma-rays.
Figure~\ref{fig:etcc} is a schematic of SMILE-2+ ETCC.
The material of a Compton-scattering target should have 
a high electron density to increase the Compton scattering probability 
and a low atomic-number to suppress the photoabsorption.
For this purpose, the electron tracker of SMILE-2+ ETCC with a sensitive volume of $30 \times 30 \times 30$~cm$^3$ 
is filled with an argon-based gas (Ar : CF$_4$ : iso C$_4$H$_{10}$; in pressure ratio = 95 : 3 : 2) 
at a pressure of 2~atm.
The approximate drift velocity of the electrons in this gas is 3.7 cm $\mu$s$^{-1}$.
As the ETCC needs three-dimensional (3D) precise electron tracks for the gamma-ray reconstruction,
we adopted a time projection chamber (TPC) with a micro pixel chamber ($\mu$-PIC) \citep{2001NIMPA.471..264O, 2005NIMPA.546..258T}
and a gas electron multiplier \citep{1997NIMPA.386..531S, 2006NIMPA.560..418T} 
insulated by 100~$\mu$m liquid crystal polymer.
To reduce the power consumption, we combined two adjacent readout-strips of $\mu$-PIC into one preamplifier.
The readout pitch of the tracker is 800~$\mu$m and
the energy resolution of the tracker through the whole volume is 45.9\% for 0.043~MeV (GdK$\alpha$) 
at full-width half maximum (FWHM).

As the gamma-ray absorber, we selected GSO (Gd$_2$SiO$_5$:Ce) pixel scintillator arrays (PSAs)
each containing $8 \times 8$~pixels.
The pixel size is $6 \times 6$~mm$^2$.
The GSO scintillator is 26 and 13~mm thick at the bottom and sides of the electron tracker, respectively.
To efficiently absorb the scattered gamma-rays,
we placed 36~PSAs at the bottom and 18~PSAs at each side of the tracker.
The total number of scintillation pixels was 6912.
For the photo readout, we adopted the 4-channel (ch) charge division method with a resistor network \citep{2006NIMPA.563...49S} 
and multi-anode photomultiplier tubes (Hamamatsu Photonics, flat-panel H8500).
The energy resolutions of the bottom and side PSAs for 0.662~MeV are 13.4\% and 10.9\%, respectively, at FWHM.
The PSAs are placed in the TPC vessel, 
whereas the scintillators of the previous ETCC (SMILE-I) were placed outside of the vessel.
The total number of readouts of TPC and PSAs are 768~ch for TPC and 432~ch, respectively.
SMILE-2+ employs the same data acquisition system as the middle-size ETCC prototype \citep{2015NIMPA.800...40M},
but with the VMe bus replaced by a gigabit Ethernet link for fast data transference.
The top of the TPC vessel is installed with
a 5~mm thick plastic scintillator to reduce the number of triggers by charged particles.

\begin{figure}
    \plotone{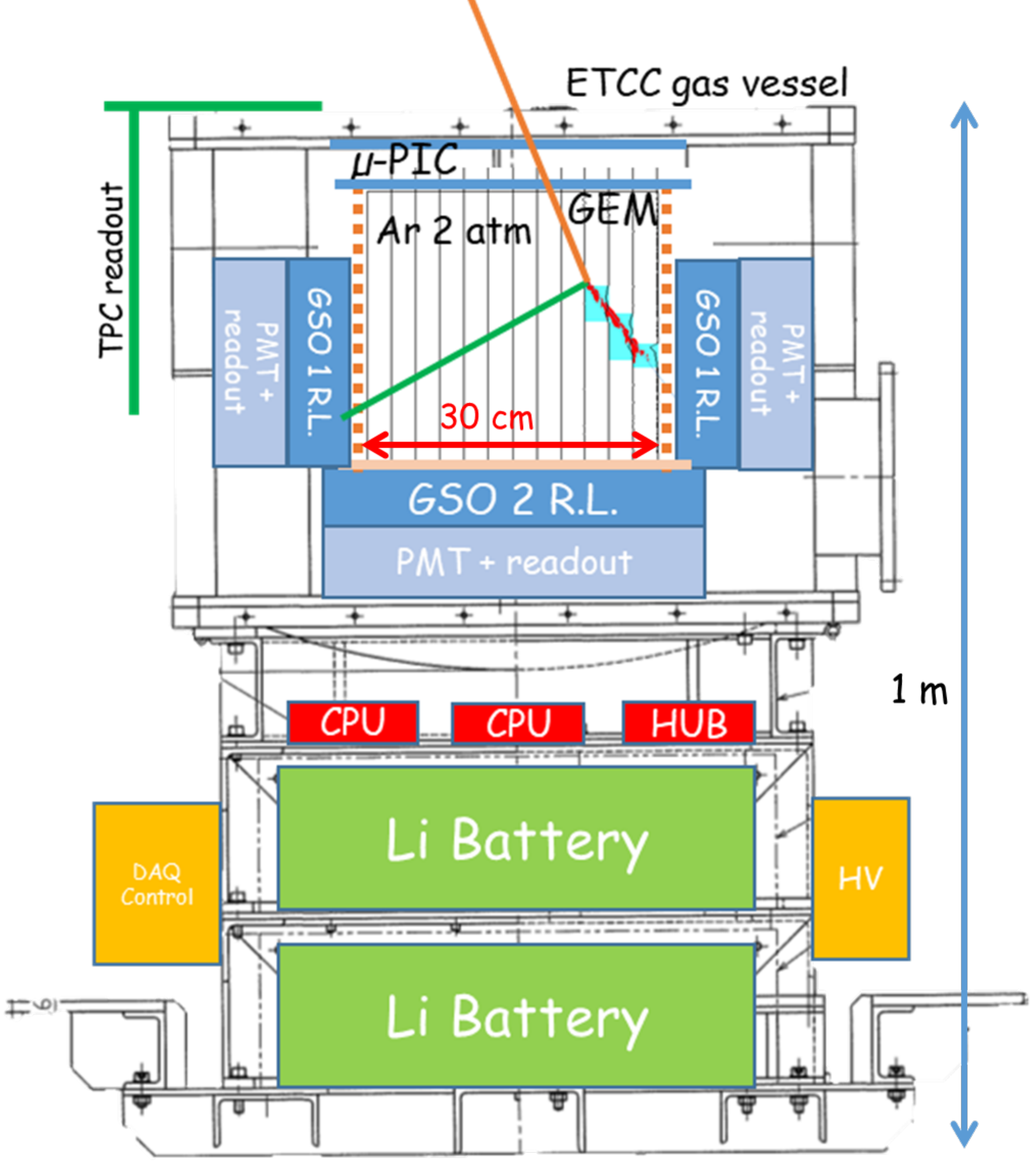}
    \plotone{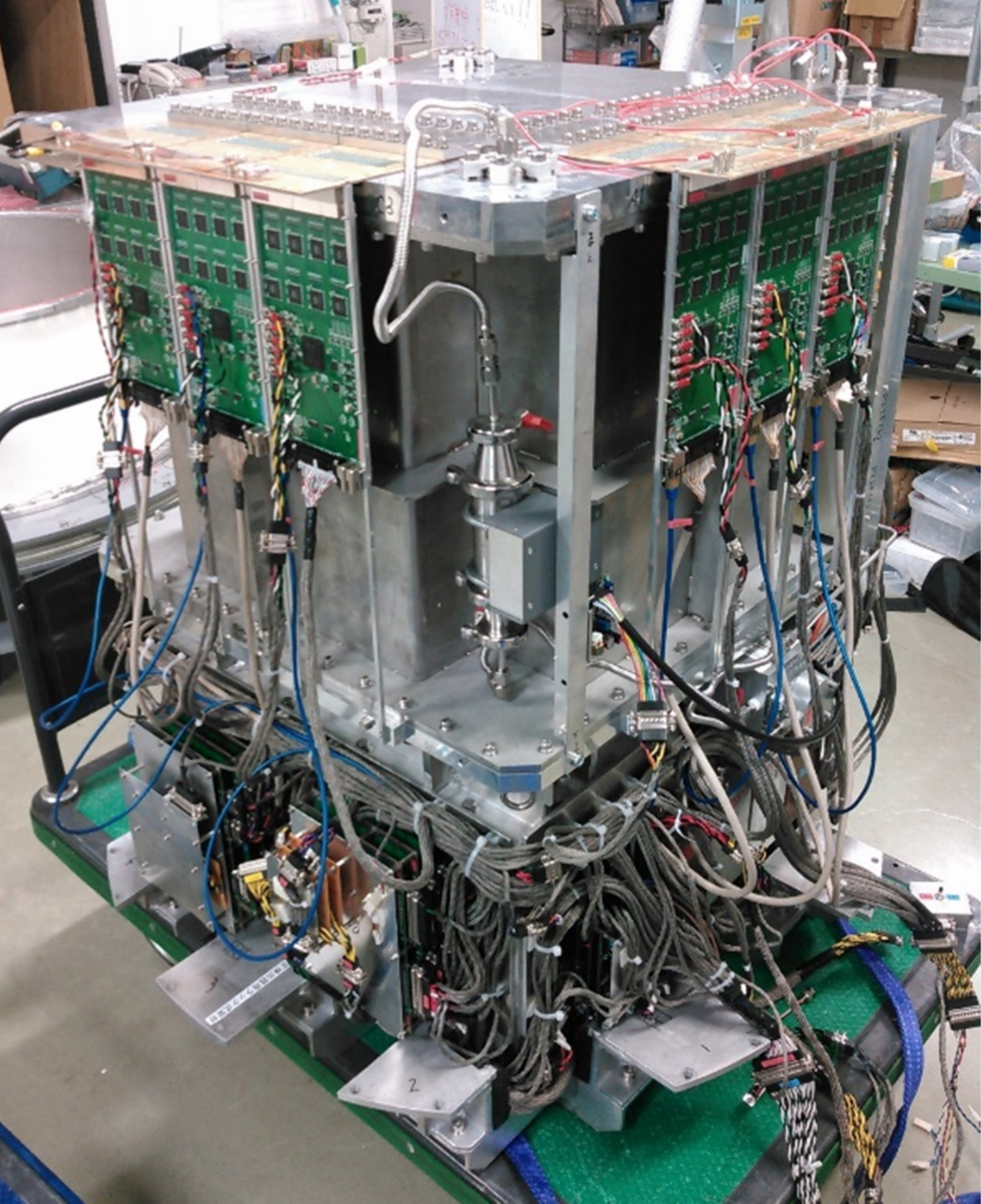}
    \caption{
	    Cross-sectional view (upper) and photograph (lower) of SMILE-2+ system.
	    Upper and lower halves are the ETCC and the control system, respectively. 
    }
	\label{fig:smile2p_etcc}
\end{figure}

\begin{figure}
    \plotone{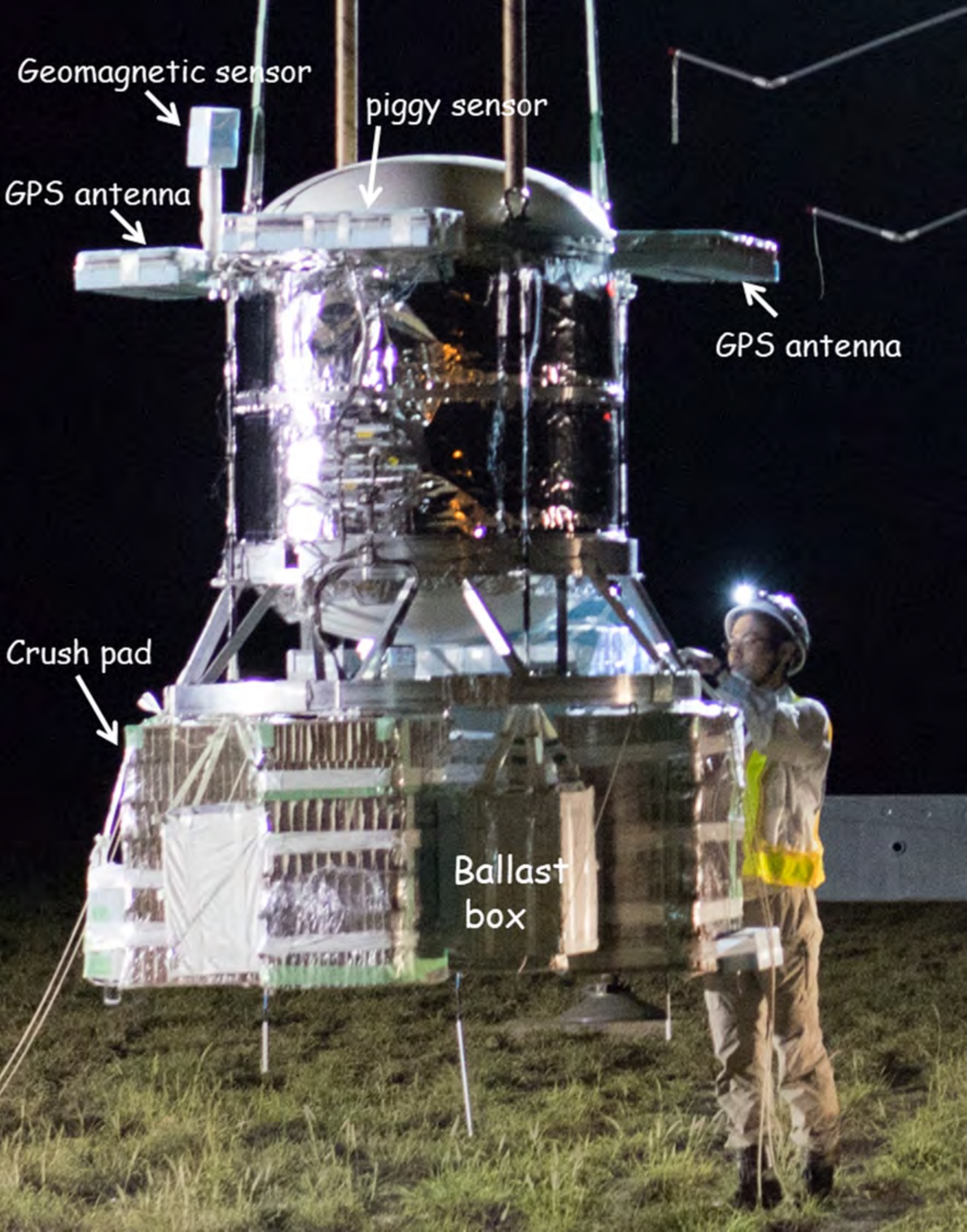}
    \caption{
	    Photograph of the SMILE-2+ gondola.
	}
	\label{fig:s2p_gondola}
\end{figure}

The SMILE-2+ ETCC is set above the control system, as shown in Fig.~\ref{fig:smile2p_etcc}.
The control system includes a central processing unit (CPU) for communication with the balloon control system, 
two CPUs with 1~TB solid-state drives for data acquisition, a trigger control unit described in \citet{2015NIMPA.800...40M},
four high-voltage units for the TPC, a power management system with DC/DC converters, and lithium batteries.
SMILE-2+ also has a receiver with a global positioning system, an atmospheric pressure gauge, two clinometers, 
and three geomagnetic aspectmeters (GAs) for measuring the gondola attitude.
However, it lacks a feedback system for attitude control.
The accuracy of posture measurements is less than 5~degrees.
The total power consumption is approximately 250~W.
Power is provided by the lithium batteries.
The SMILE-2+ system is sealed in a pressured vessel maintained at 1~atm.
The side of the outer vessel is covered by multilayered insulators for temperature maintenance,
and the outer vessel is placed on the small aluminum gondola, as shown in Fig.~\ref{fig:s2p_gondola}.
The outer vessel is installed with an independent piggyback sensor~\citep{2019ISTS_Shoji} 
that measures the attitude with three GAs, three accelerometers, and a gyroscope. 
The gondola attitudes determinanted by the SMILE-2+ sensors and the piggyback sensor
were checked for consistency.
Without ballast, the SMILE-2+ gondola weights 511~kg in total.

\subsection{ground calibration}\label{sec:cal}
The gaseous electron tracker on the ETCC obtains the 3D tracks 
and energies of the Compton-recoil electrons,
whereas the absorber detects the absorption points and energies of the Compton-scattered gamma-rays.
The momentum of the incident gamma-rays is then obtained
by summing the momenta of the recoil electrons and the scattered gamma-rays:
\begin{equation}
	p_0^\mu = p_\gamma^\mu + p_e^\mu ,
\end{equation}
where $p_0^\mu$, $p_\gamma^\mu$, and $p_e^\mu$ are the four-dimensional momenta of the incident gamma-ray,
scattered gamma-ray, and recoil electron, respectively.
Using the measured values, the unit vector of the incident gamma-ray ${\bf r}$ is described by
\begin{equation}
	{\bf r}
		= \left( \cos\phi  - \frac{\sin  \phi}{\tan \alpha}  \right) {\bf g}
		+ \frac{\sin \phi}{\sin \alpha} {\bf e} ,
\end{equation}
where $\bf g$ and $\bf e$ are unit vectors in the directions 
of the scattered gamma-ray and the recoil electron, respectively, in the laboratory system.
$\alpha$ is the angle between the scattering and recoil directions (see Fig.~\ref{fig:etcc}),
and $\phi$ is the scattering angle given by
\begin{equation}
    \label{eq:scattering_angle}
	\cos \phi = 1 - m_e c^2 \left( \frac{1}{E_\gamma} - \frac{1}{E_\gamma + K_e} \right) .
\end{equation}
In Eq.~(\ref{eq:scattering_angle}), 
$E_\gamma$, $K_e$, $m_e$, and $c$ denote the energy of the scattered gamma-rays, 
kinetic energy of the Compton-recoil electrons, the electron mass, and light speed, respectively.

\begin{figure}
    \plotone{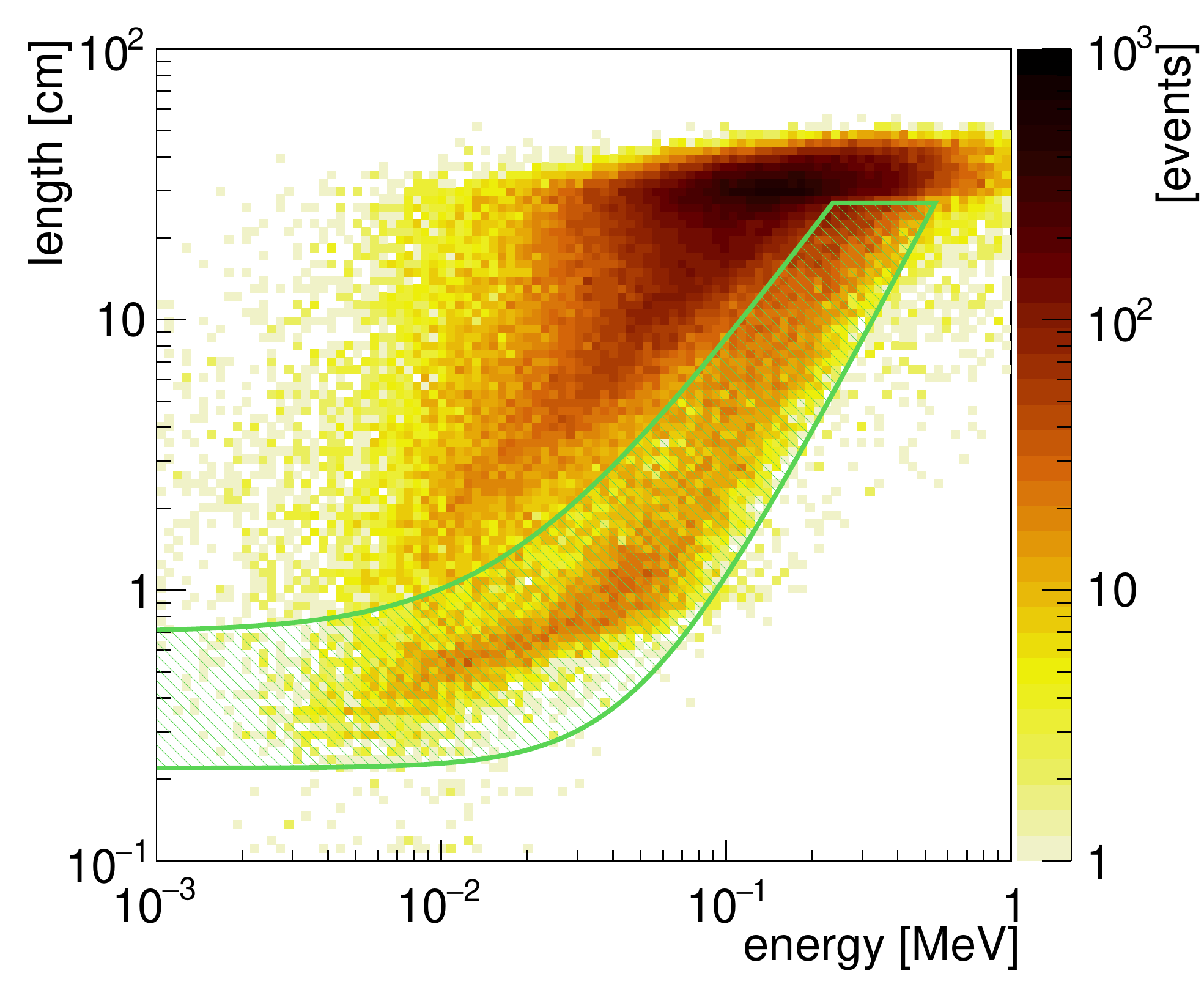}
    \caption{
	    Track length of charged particles as a function of deposited energy in SMILE-2+ TPC.
	    We selected only the fully-contained electron events inside the hatched green area.
    }
	\label{fig:dedx}
\end{figure}

The gamma-ray candidate events are reconstructed under the following criteria;
\begin{enumerate}
	\item Single pixel scintillator hits:
		When a Compton-scattered gamma-ray hits more than one pixel in the absorber,
		the incident gamma-ray is difficult to reconstruct
		because the sequence of interactions in the absorber become confused.
		Therefore we select only the events with a single pixel scintillator hit.
	\item Fully-contained electrons:
		The gamma-ray reconstruction requires the kinetic energy of a recoil electron. 
		If a recoil electron escapes the sensitive volume of the TPC, 
		the incident gamma-ray cannot be reconstructed 
		because the recoil energy measurement is incomplete.
		We thus set the TPC fiducial volume at 29$\times$29$\times$29~cm$^3$
		and require that the track length matches 
		the expected distance range of electrons depositing their energy 
		in Ar gas \citep{2015ApJ...810...28T}.
		Figure~\ref{fig:dedx} plots the track length of the charged particles detected 
		in level flight versus the energy deposited in the TPC.
		The gradient in this figure represents the energy loss $dE/dx$.
		The events in the hatched area of this figure give rise to fully-contained electrons.
		The head-tail of the recoil electrons is determined 
		from the skewness of the track image \citep{2008NIMPA.584..327D}
		and the recoil direction determined from the time-over-threshold information \citep{2015ApJ...810...28T}.
		The angular resolution of the recoil direction and the position resolution of the scattering points 
		are determined by a traditional method (see \citet{2021arXiv210502512I}).
	\item Compton scattering kinematics:
		$\alpha$ is defined as
		\begin{equation}
			\cos \alpha_g = {\bf g}  \cdot  {\bf e},
		\end{equation}
		and can be calculated by Compton-scattering kinematics as follows: 
		\begin{equation}
			\cos \alpha_k = \left( 1 - \frac{m_e c^2}{E_\gamma} \right) \sqrt{\frac{K_e}{K_e + 2m_e c^2}}.
		\end{equation}
		Therefore, we can select only Compton scattering events with the condition described by
		\begin{equation}
			\left| \cos \alpha_g - \cos \alpha_k \right| \leq \Delta_\alpha ,
		\end{equation}
		where $\Delta_\alpha$ is a cut parameter.
		For SMILE-2+ ETCC, we set $\Delta_\alpha = 0.5$.
\end{enumerate}

\begin{figure}
    \plotone{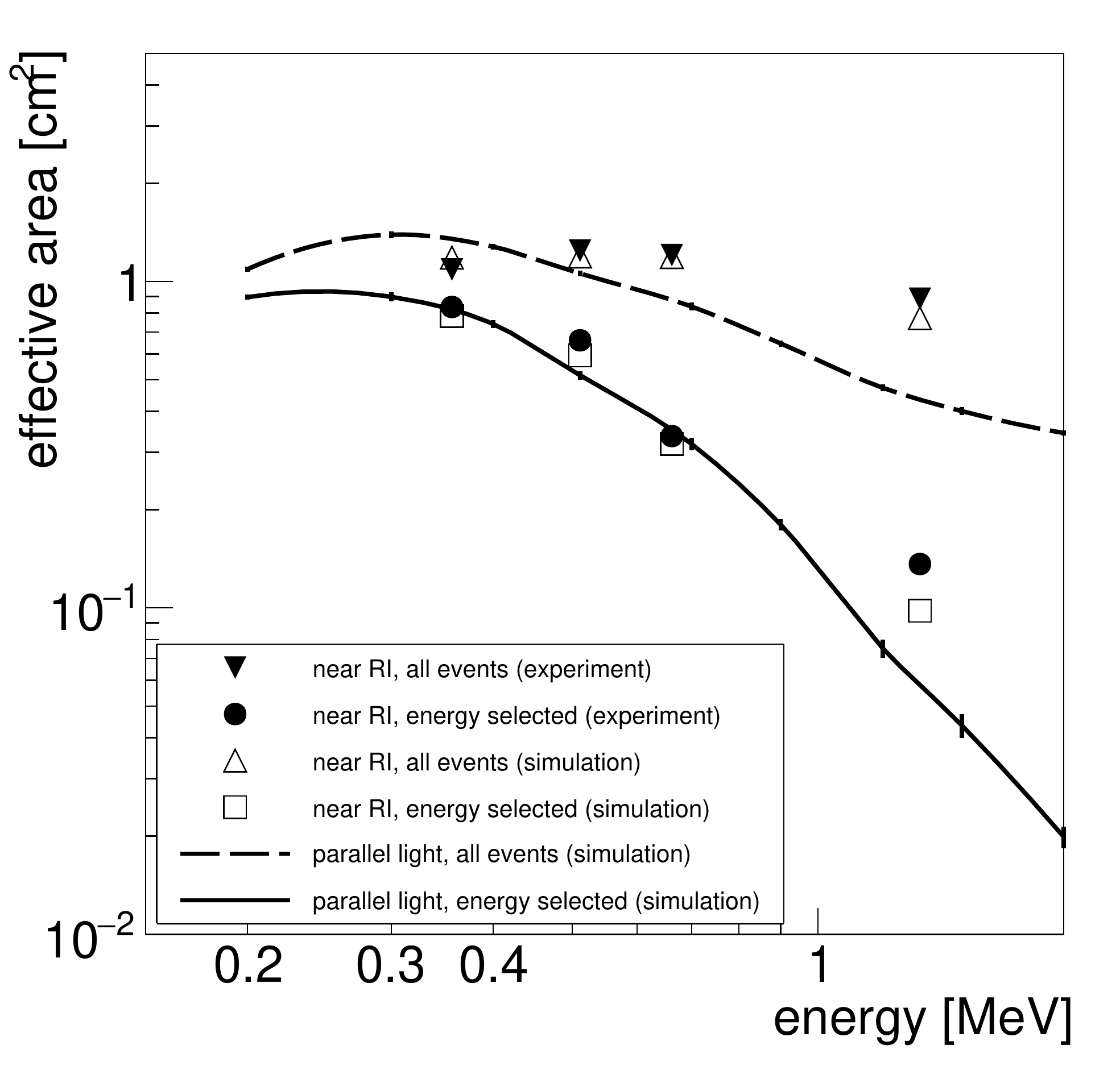}
    \caption{
	    Effective areas as functions of incident energy.
	    Filled and open triangles represent the effective areas of all reconstructed events 
	    obtained in the experiments and simulation, respectively.
	    Filled circles and open squares represent the experimental and simulated effective areas
	    after selection within the FWHM of the energy peak, respectively.
	    Dashed and solid lines plot the effective areas of all events and energy-selected events
	    in the parallel light case, respectively.
    }
	\label{fig:area}
\end{figure}

\begin{figure}
    \plotone{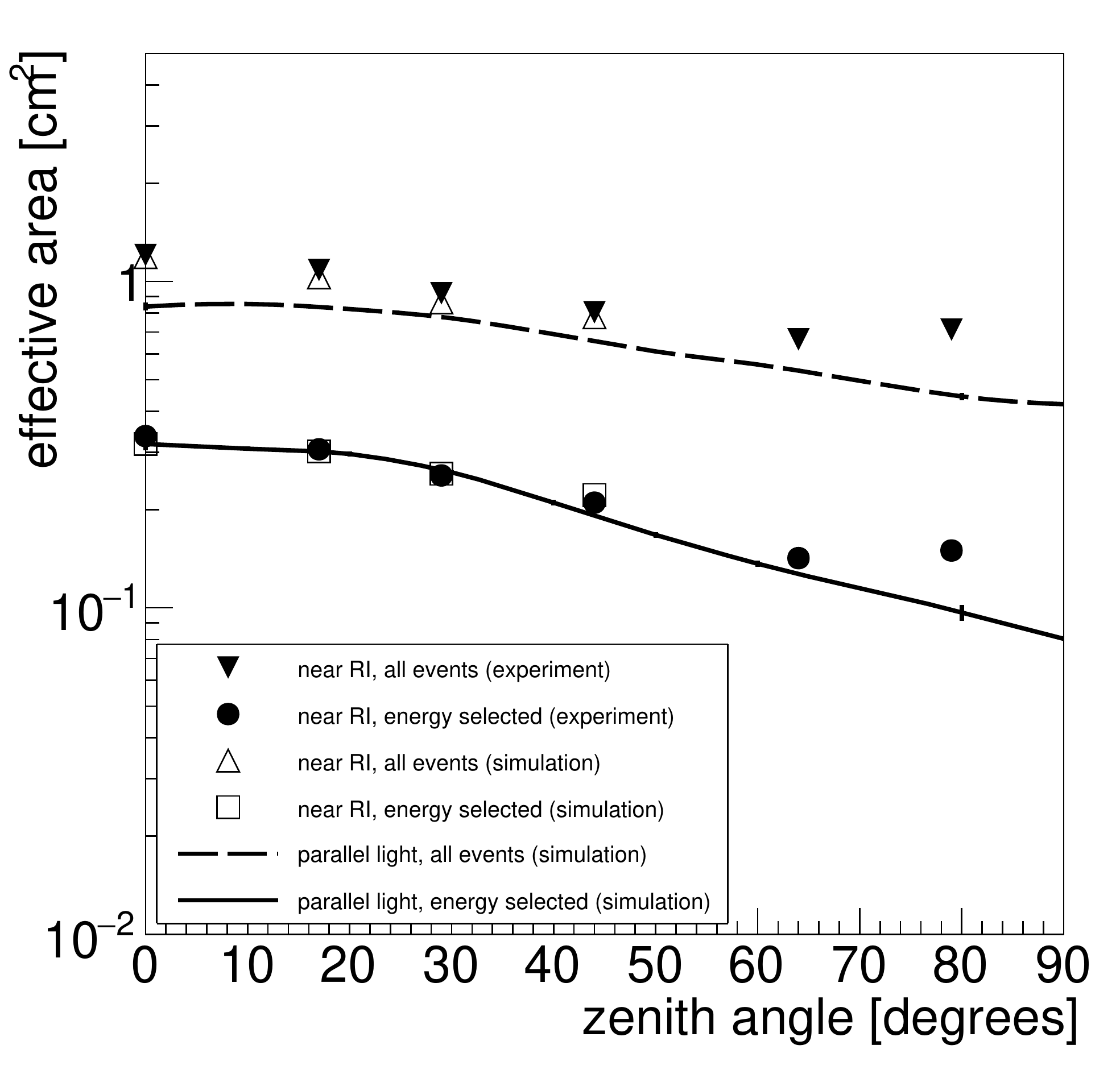}
    \caption{
	    Zenith angle dependence of the effective area at an incident energy of 0.662~MeV.
	    Symbols are described in the caption of Fig.~\ref{fig:area}.
	}
	\label{fig:zenith}
\end{figure}

\begin{figure}
    \plotone{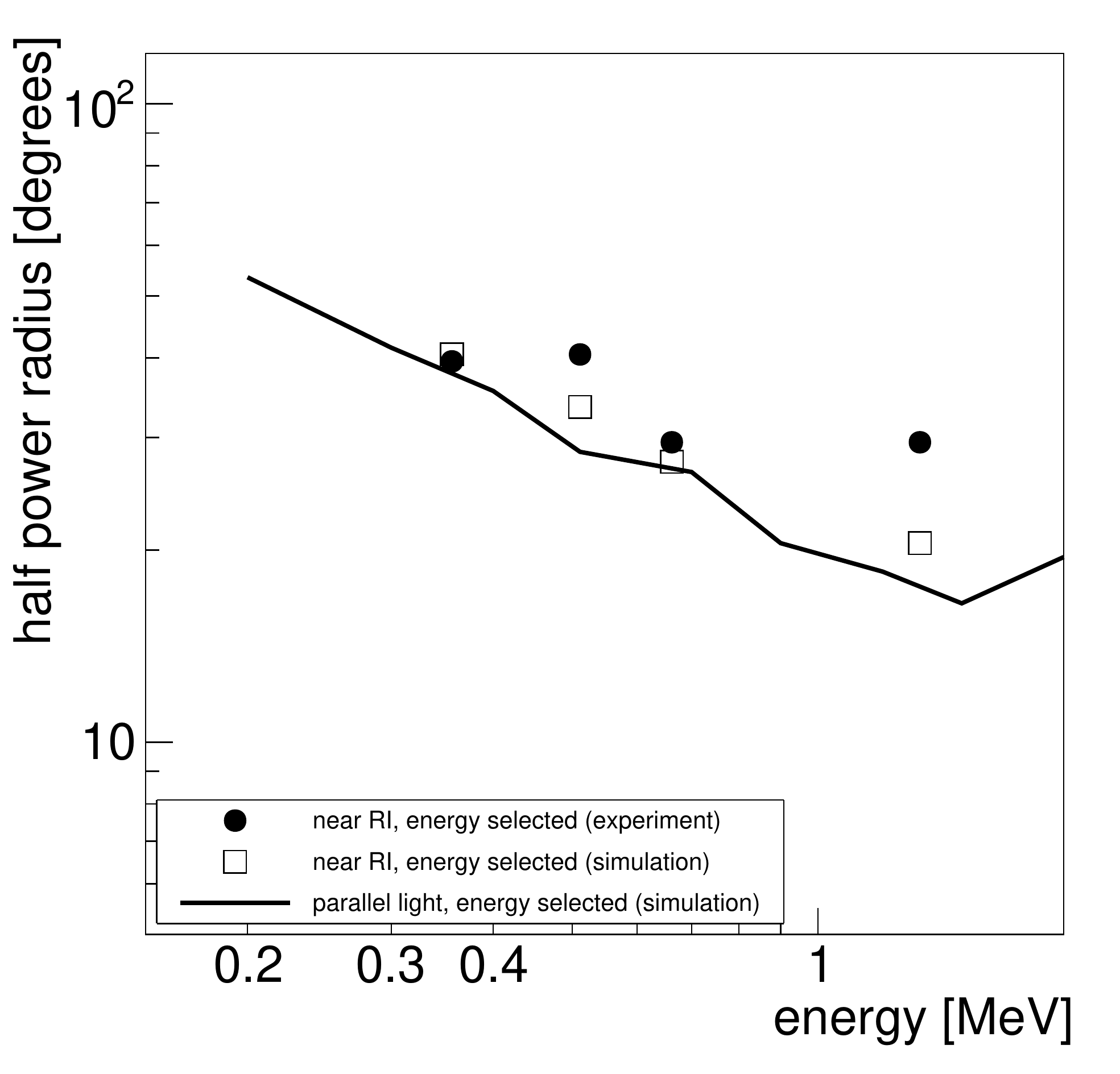}
    \caption{
	    Half power radius (HPR) of the PSF as a function of incident energy.
	    The filled circles, open squares, and solid line represent the measured HPR using the checking sources,
	    the simulated HPR with a near point source, and the simulated HPR for parallel light, respectively.
    }
	\label{fig:psf}
\end{figure}

\begin{figure}
    \plotone{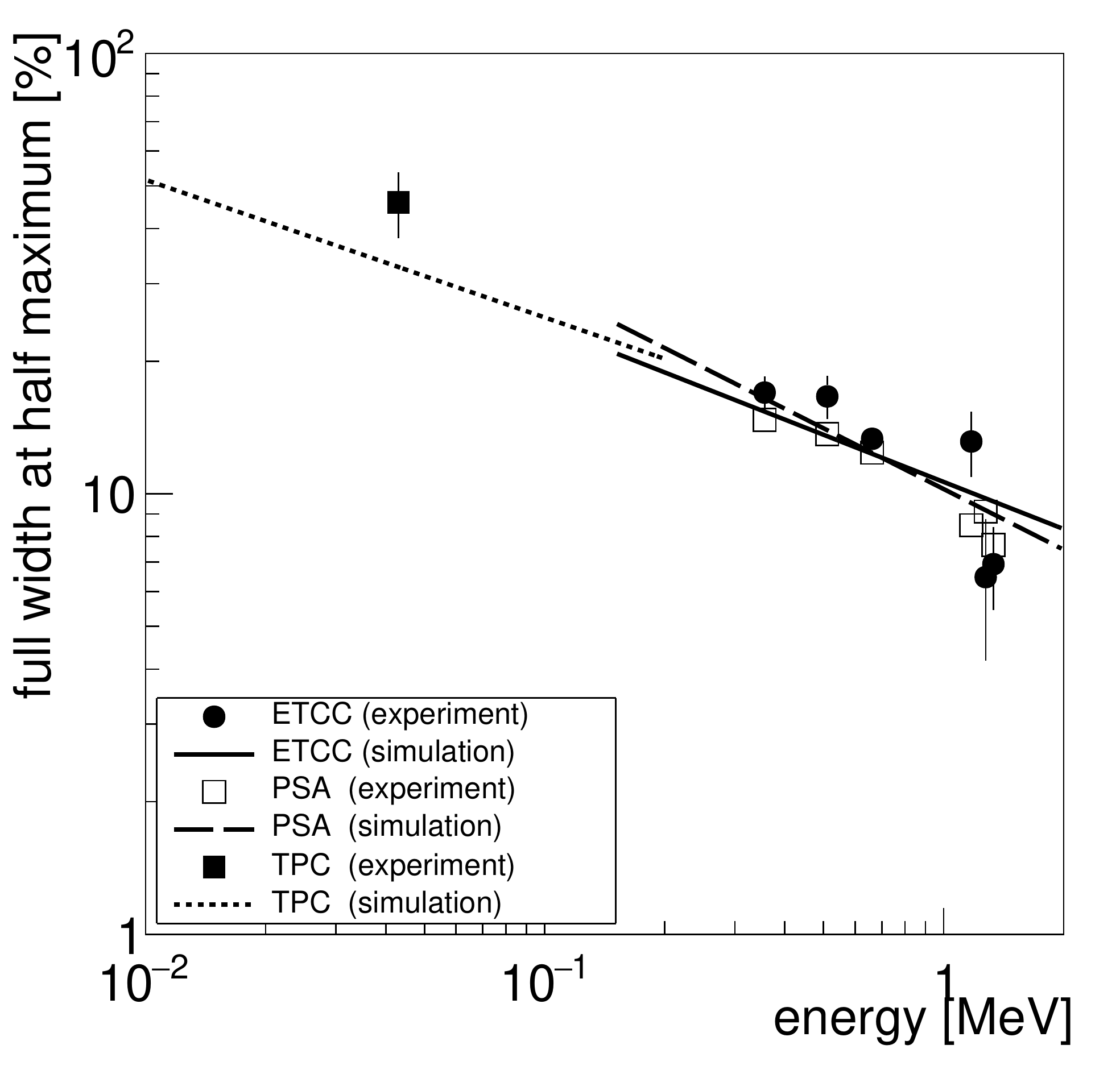}
    \caption{
	    Energy resolutions of the ETCC, TPC, and PSAs as functions of energy.
	    The filled circles and solid line represent the energy resolutions of the SMILE-2+ ETCC obtained 
	    via ground calibration and simulation, respectively.
	    The filled and open squares are the averaged energy resolutions of the TPC and PSAs, respectively,
	    measured via ground calibration.
	    The dotted and dashed lines represent the simulated energy resolutions of TPC and PSA, respectively.
    }
	\label{fig:E_res}
\end{figure}

The response function of parallel light,
which is required for deconvoluting the gamma-ray fluxes from celestial objects,
was obtained in a simulator of SMILE-2+ ETCC based on Geant4 \citep[ver 10.04-patch02;][]{2003NIMPA.506..250A}.
Electromagnetic interactions were calculated in G4EmLivermorePhysics
while considering the Doppler broadening effect of Compton scattering.
To confirm the reliability of the SMILE-2+ ETCC simulator, 
we measured the effective area, PSF, and energy resolution 
under irradiation by line gamma-rays from the checking sources
placed approximately 2~m from the center of SMILE-2+ ETCC.
The measured performances were compared with the simulated expectations.
Figure~\ref{fig:area} plots the effective areas as functions of incident energy
when the checking sources were placed along on the center axis of the ETCC 
or when parallel light was irradiated at a zenith angle of 0~degrees.
The expected and measured effective areas were consistent. 
In the energy-selected results, the realized effective area was 1.1~cm$^2$ at 0.356~MeV, 
which satisfies the criterion for detecting the Crab nebula.
The difference between all reconstructed events 
and the energy-selected events (twice the FWHM of the full-energy peak) increased at the higher energies.
This difference is caused by the scattered component,
namely, the scattered gamma-rays in the surrounding materials 
(e.g., the pressured vessel, TPC vessel, and PSA support structures) 
before the radiation enters the ETCC.
As the major interaction between incoming rays and materials is Compton scattering,
observations are considered to be confused 
not only by PSF blurring of the surrounding sources (the expected contamination)
but also by components scattered from the structures.
Therefore, an accurate response function is critical for obtaining the true fluxes of celestial objects.
The zenith angle dependence of the effective area at 0.662 MeV is shown in Fig.~\ref{fig:zenith}.
As mentioned above,
SMILE-2+ ETCC has a large FoV (3.1 sr).
This FoV is defined as the field covering more than half the effective area at the zenith.
In the ETCC simulator,
the accuracy of the effective area was better than 10\% 
over the energy range 0.15--2.1~MeV within the FoV.
Figure~\ref{fig:psf} plots the Half-power radius (HPR) of PSF as a function of incident energy.
The PSF size of SMILE-2+ ETCC is 30 degrees at the HPR for 0.662~MeV.
Furthermore, the angular resolution measure (ARM) 
and scatter plane deviation (SPD) at the FWHM for 0.662~MeV,
which define the accuracies of the scattering angle and the scattering plane, respectively,
are 10.5 and 148~degrees for 0.662~MeV at FWHM.
The PSF of the ETCC depends on the energy resolution of the PSAs, 
the position accuracy of the Compton-scattering point, 
and the angular resolution of the direction of the Compton-recoil electrons.
Although SMILE-2+ ETCC has poorer spatial resolution than the advanced telescopes that observe other wavelengths,
it satisfies the criteria for detecting the Crab nebula
with powerful background rejection capability.
Figure~\ref{fig:E_res} shows the energy resolutions of SMILE-2+ ETCC, TPC and PSA.
The fully-contained electron events are limited to recoil energies lower than 0.3~MeV 
because the TPC gas lacks any stopping power.
Therefore the energy of the scattered gamma-ray exceeds that of the recoil electron 
and the energy resolution of the ETCC is dominated by that of PSAs.

\section{SMILE-2+ Balloon flight}
\begin{figure}
    \plotone{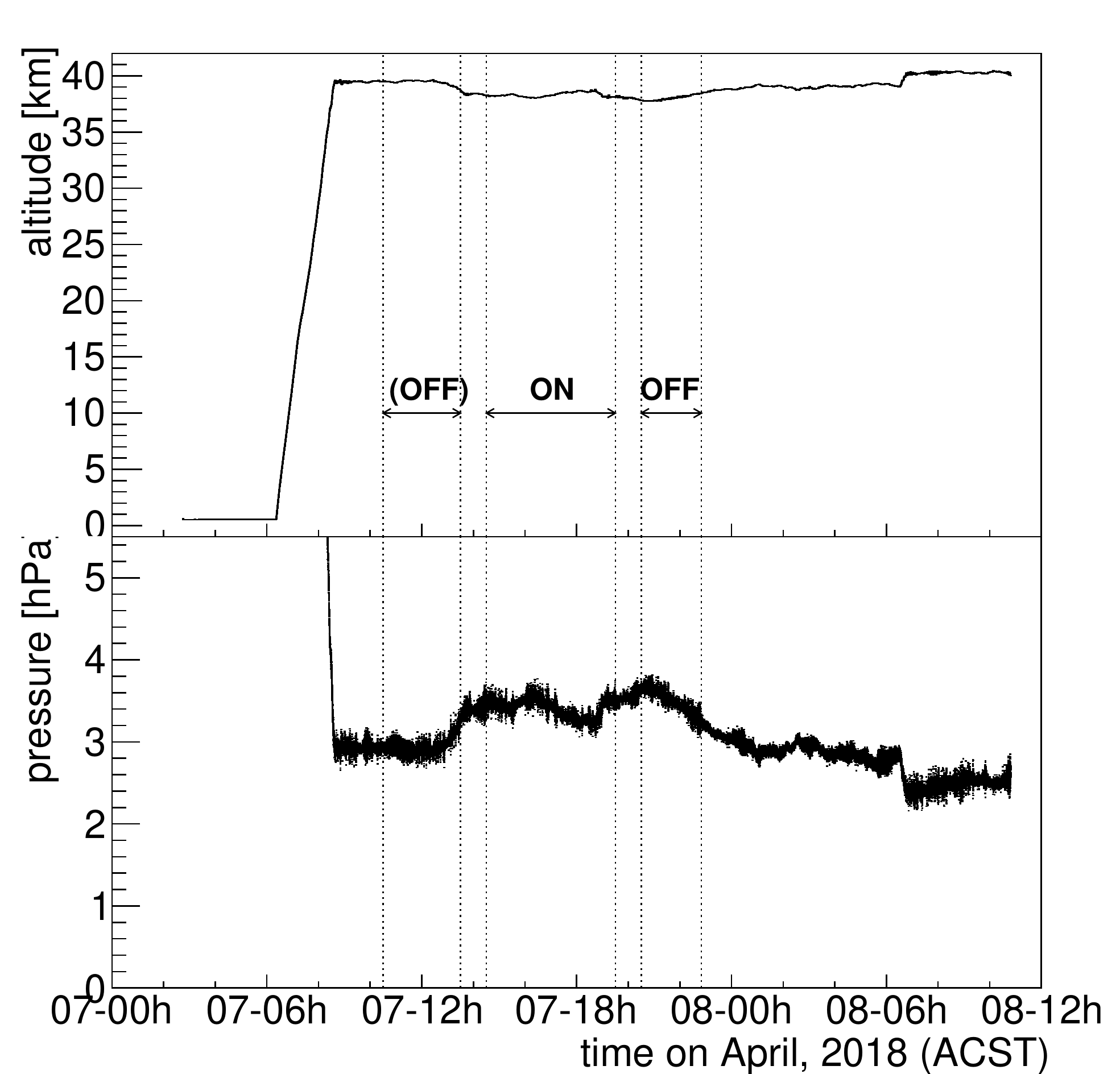}
    \caption{
	    Time variations of the altitude of SMILE-2+ gondola (upper panel) and atmospheric pressure (lower panel).
        During the 'ON' period, Crab nebula was included in the FoV and was observed.
        'OFF' and '(OFF)' denote the period of the background modeling 
        when the atmospheric depth was the same and thinner 
        than that of during the ON periods, respectively.
        During those periods, the Crab nebula was excluded from FoV of SMILE-2+.
    }
	\label{fig:altitude}
\end{figure}

The SMILE-2+ balloon was launched by ISAS/JAXA from the Australian balloon launch station, 
Alice Springs, Australia, on April 7 of 2018.
Three hours before launching at 06:24 Australian Central Standard Time (ACST),
the SMILE-2+ system was switched on the data acquisition was running even during the ascent.
The time variations of the altitude and atmospheric pressure are shown 
in the upper and lower panels of Fig.~\ref{fig:altitude}, respectively.
At 08:44 ACST, the balloon reached an altitude of 39.6~km.
SMILE-2+ performed observations until 10:45 ACST on April 8 of 2018, and was switched off at 10:53 ACST.
At 06:30 ACST on April 8 of 2018, the balloon slightly ascended owing to sunrise.
The duration of the level flight, in which the atmospheric depth 
was maintained between 2.4--3.8~hPa (altitude 37.8--40.4~km), was approximately 26~h.
On April~9 of 2018, we approached and successfully recovered the SMILE-2+ gondola,
which had landed approximately 190~km from Alice Springs.

\begin{figure}
    \plotone{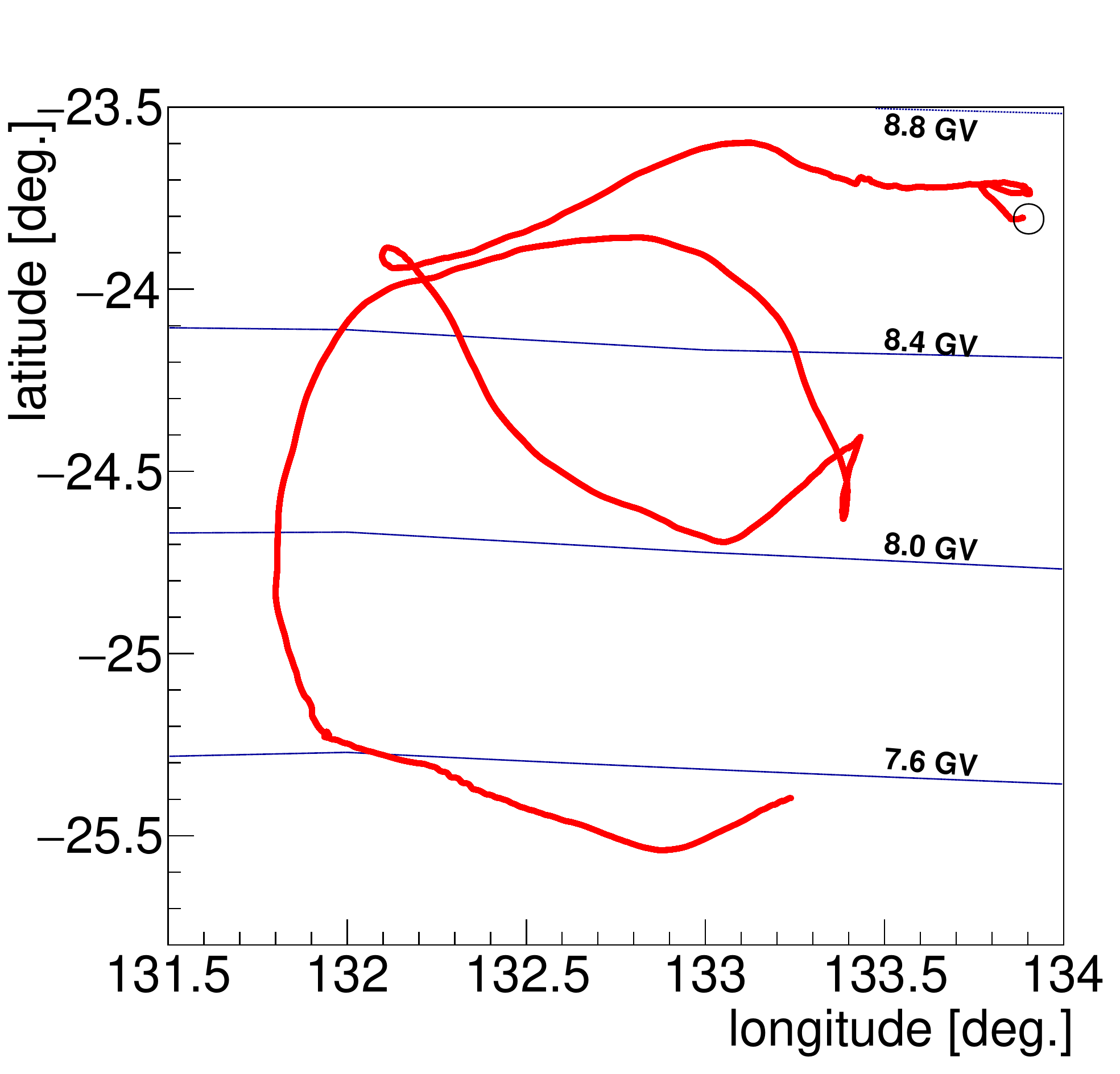}
    \caption{
	    Flight path of SMILE-2+.
	    Open circle represents the position of the Australian balloon launch station,
	    and the contours outline the cutoff rigidity calculated by PARMA \citep{2008RadR..170..244S}.
    }
	\label{fig:path}
\end{figure}

Figure~\ref{fig:path} shows the flight path of the SMILE-2+ balloon 
from the launch time until turn-off of the SMILE-2+ system.
The cutoff rigidity (marked by the contours in Fig.~\ref{fig:path}) were calculated 
by PARMA \citep{2008RadR..170..244S} based on MAGNETOCOSMICS \citep{2005IJMPA..20.6802D}.
The time-averaged cutoff rigidity was 8.2$\pm$0.4~GV during the level flight,
and the $K_p$ index, which indicades an indicator of the disturbances in the Earth's magnetic field, 
was below 2\footnote{https://www.gfz-potsdam.de/en/kp-index}.
The low $K_p$ index confirmed a stable and quiescent magnetic field condition,
under which the intensities of cosmic rays should cause negligible fluctuations in the SMILE-2+ observation.
In contrast, 
the intensity of atmospheric gamma-rays is 
proportional to $z R_{\mathrm{cut}}^{-1.13}$ \citep{1977ApJ...217..306S, 1981JGR....86.1265T},
where $z$ and $R_{\mathrm{cut}}$ are the atmospheric depth and the cutoff rigidity, respectively.
Between 13:32 and 23:00 ACST on April 7 of 2018 (see Fig.~\ref{fig:altitude}), 
the balloon altitude decreased with a concomitant 20\% increase in atmospheric depth.
The intensity of atmospheric gamma-ray was expected to rise at this time.
\begin{figure}
	\plotone{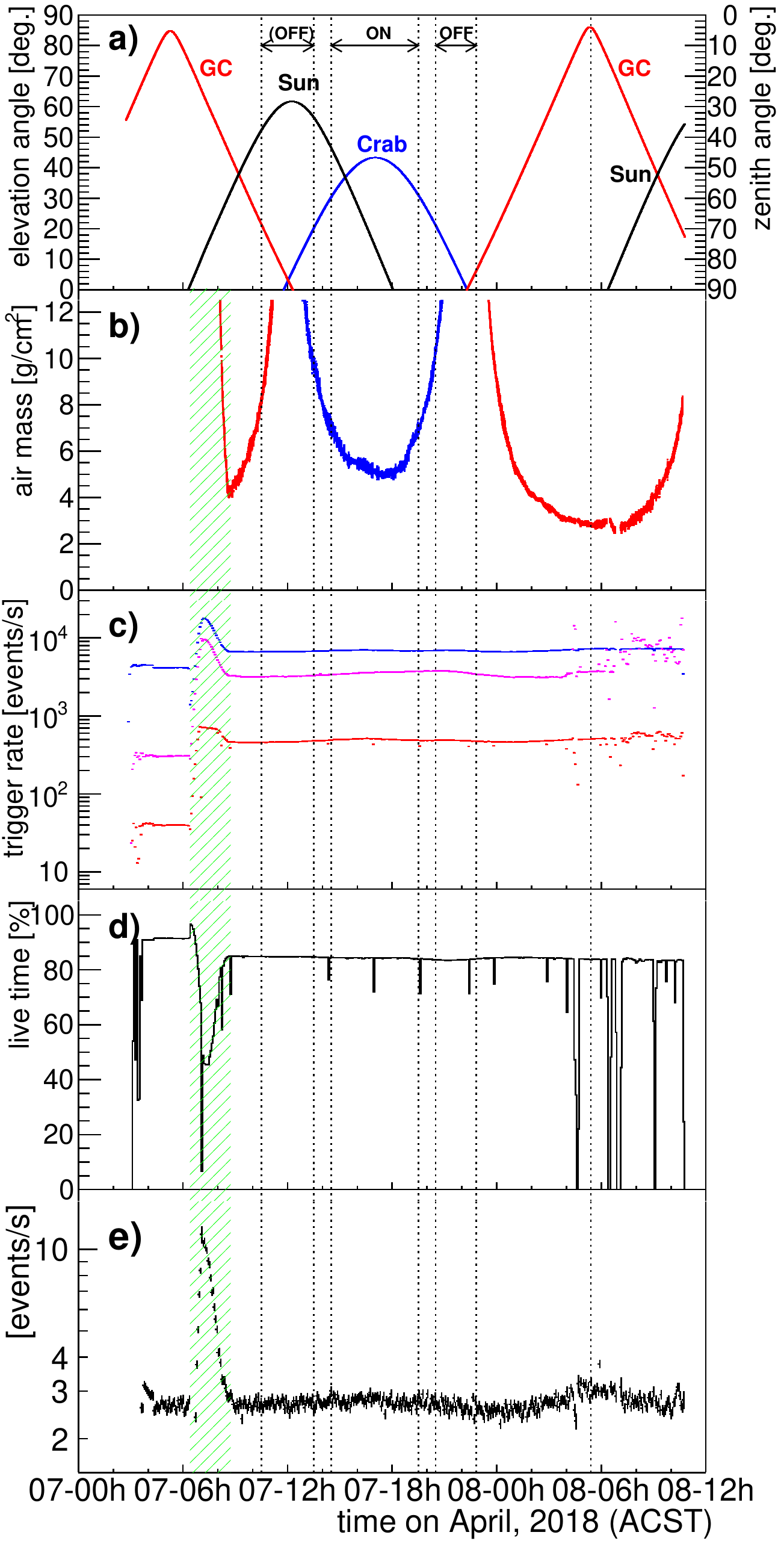}
	\caption{
		Time variations of a) elevation angles and b) air mass during the SMILE-2+ experiment.
		Red, blue, and black represent the Galactic center, the Crab nebula, and the Sun, respectively.
		c) Count rates of PSAs (blue), TPC (magenta), and ETCC (red) as functions of time;
		d) live time of the data acquisition;
		e) light curve obtained after gamma-ray reconstruction and live-time correction.
		This light curve is the total event rate of the final remaining gamma-rays in all direction.
		The hatched area in b), c), d), and e) represents the balloon ascending period.
	}
    \label{fig:air_mass}
\end{figure}

\begin{figure}
    \plotone{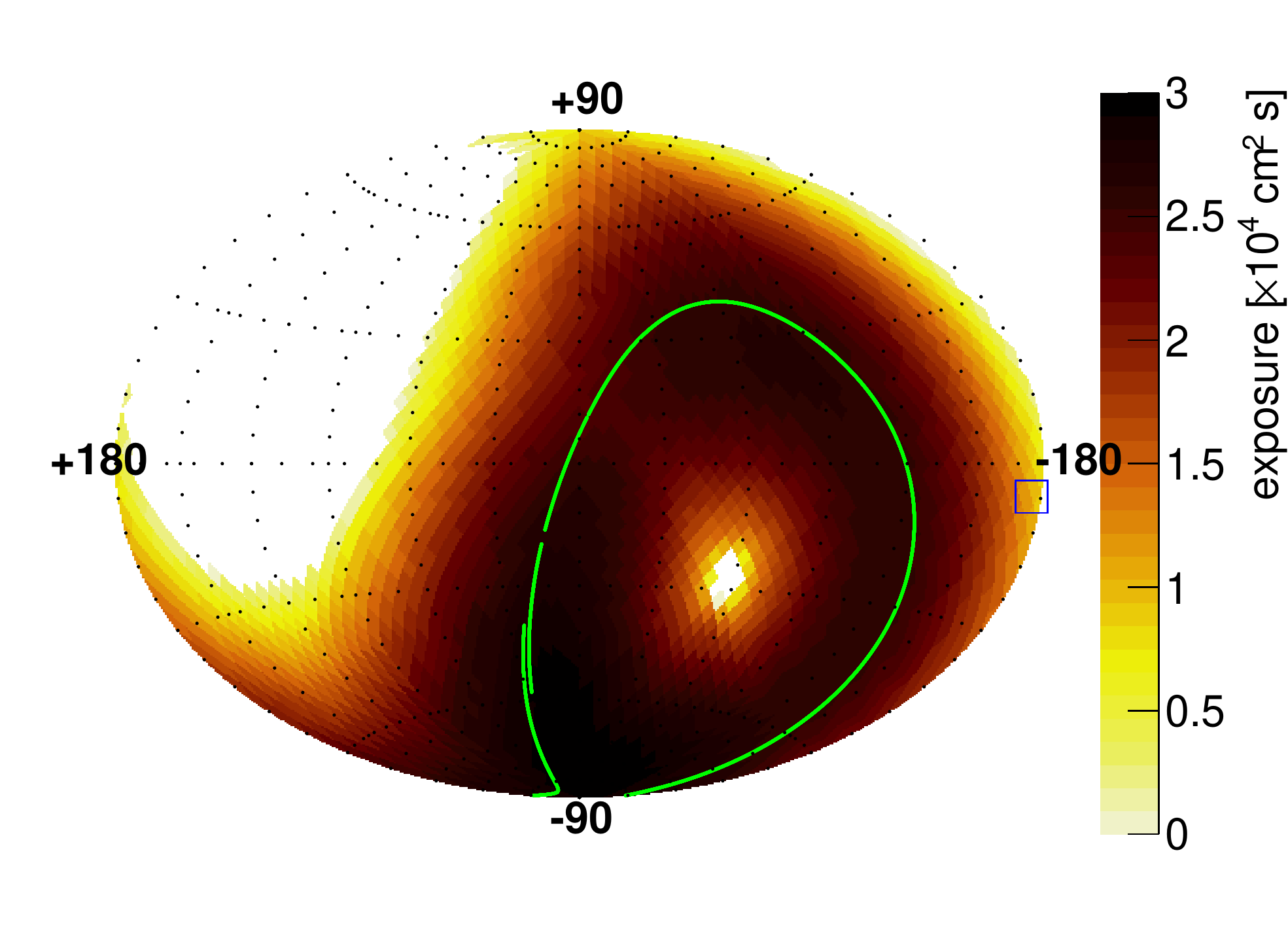}
    \caption{
 	    Exposure map of 0.3~MeV at zenith angles less than 60~degrees
 	    (in the galactic coordinates).
 	    The solid line represents the tracked light axis of SMILE-2+ ETCC 
 	    and the open square indicates the Crab nebula.
    }
 	\label{fig:map}
\end{figure}

The time variations of the elevation angles and air masses of the observation targets 
are shown in panels a) and b) of Fig.~\ref{fig:air_mass}, respectively.
Panel c) of this figure shows the count rates of PSAs, TPC, and ETCC.
The live times of SMILE-2+ 
observing the Crab nebula and the Galactic center were 5.1~h and 10.2~h, respectively.
The data acquisition rate (red plots in Fig.~\ref{fig:air_mass}) was 40~events~s$^{-1}$ on the ground,
700~events~s$^{-1}$ near the Pfotzer maximum (live time 45\%), and approximately 450~events~s$^{-1}$  at 39.6~km altitude.
After 06:45 ACST on April~8 of 2018, the count rate of the TPC was raised 
by small discharges occuring in the TPC.
However, as these discharge events can be clearly discriminated 
from the charged particle tracks in the track images, 
they do not disturb the gamma-ray observations.
During level flight, the live time of the data-acquisition was maintained above 82\% (Fig.~\ref{fig:air_mass} d)),
sufficient for observing the Crab nebula and Galactic center.
Figure~\ref{fig:map} represents the exposure map 
with the definition of the observation area by the zenith angle below 60~degrees.

\section{Analysis}
\subsection{gamma-ray reconstruction}
\begin{figure}
    \plottwo{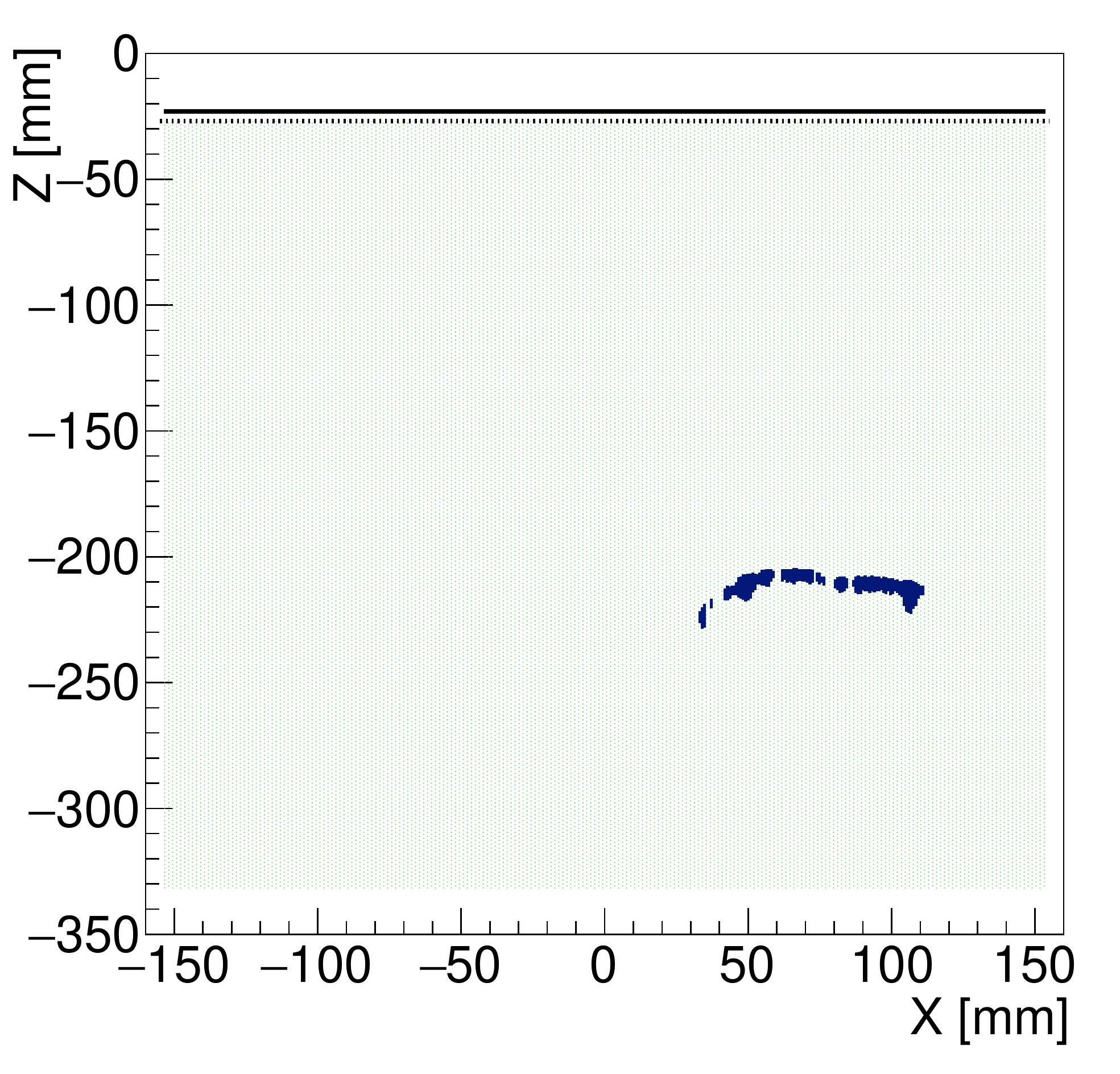}{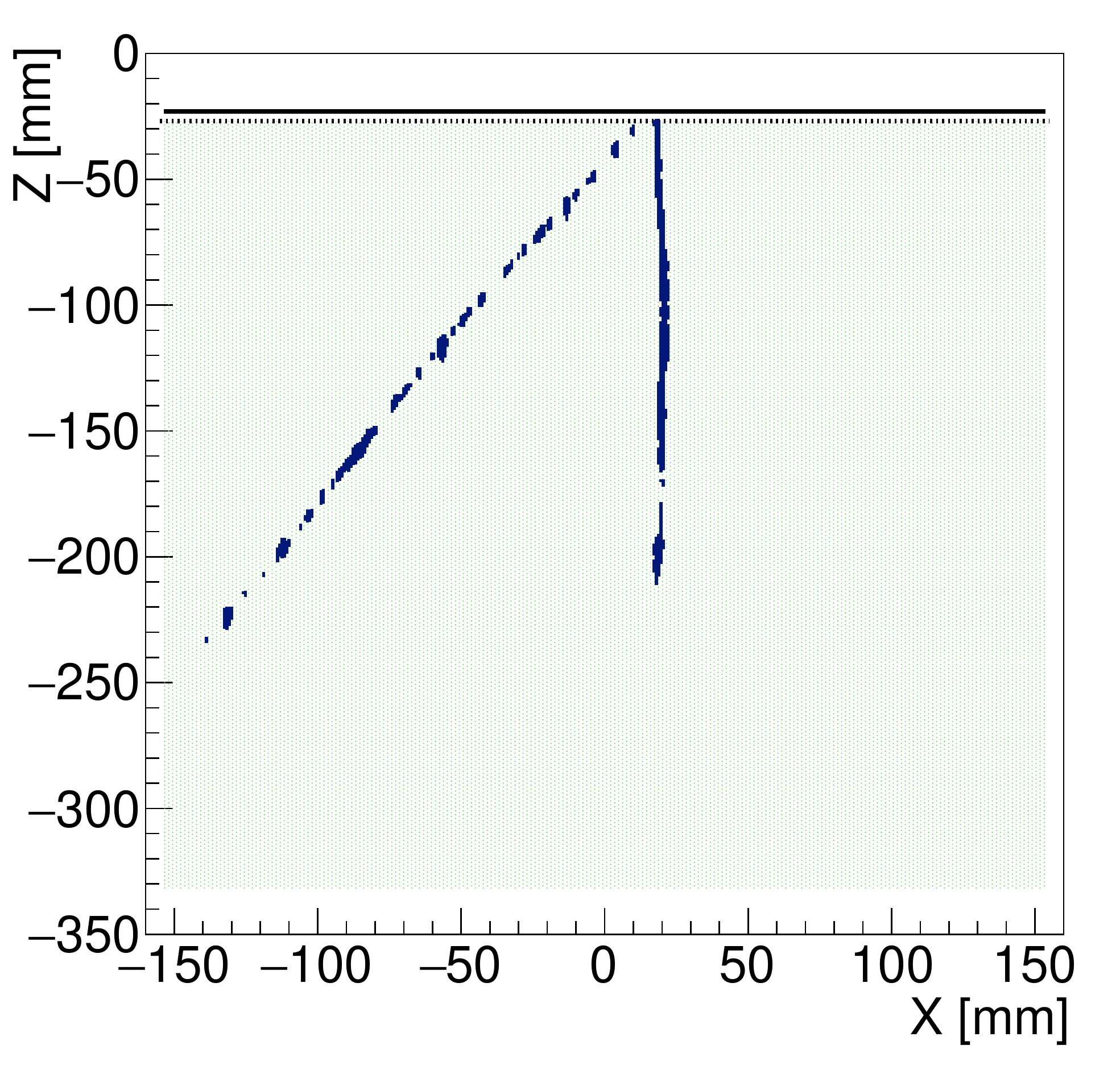}
    \plottwo{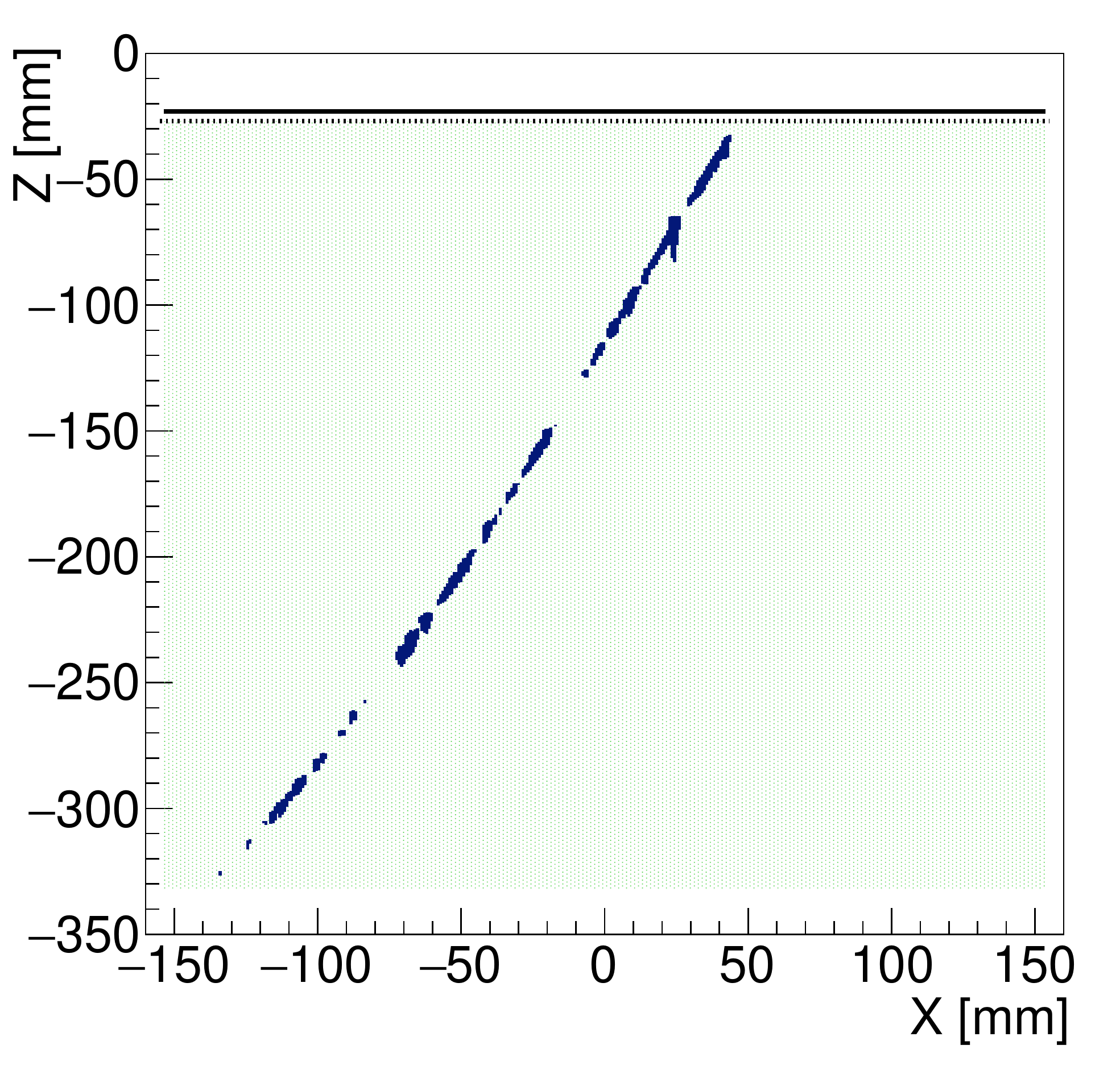}{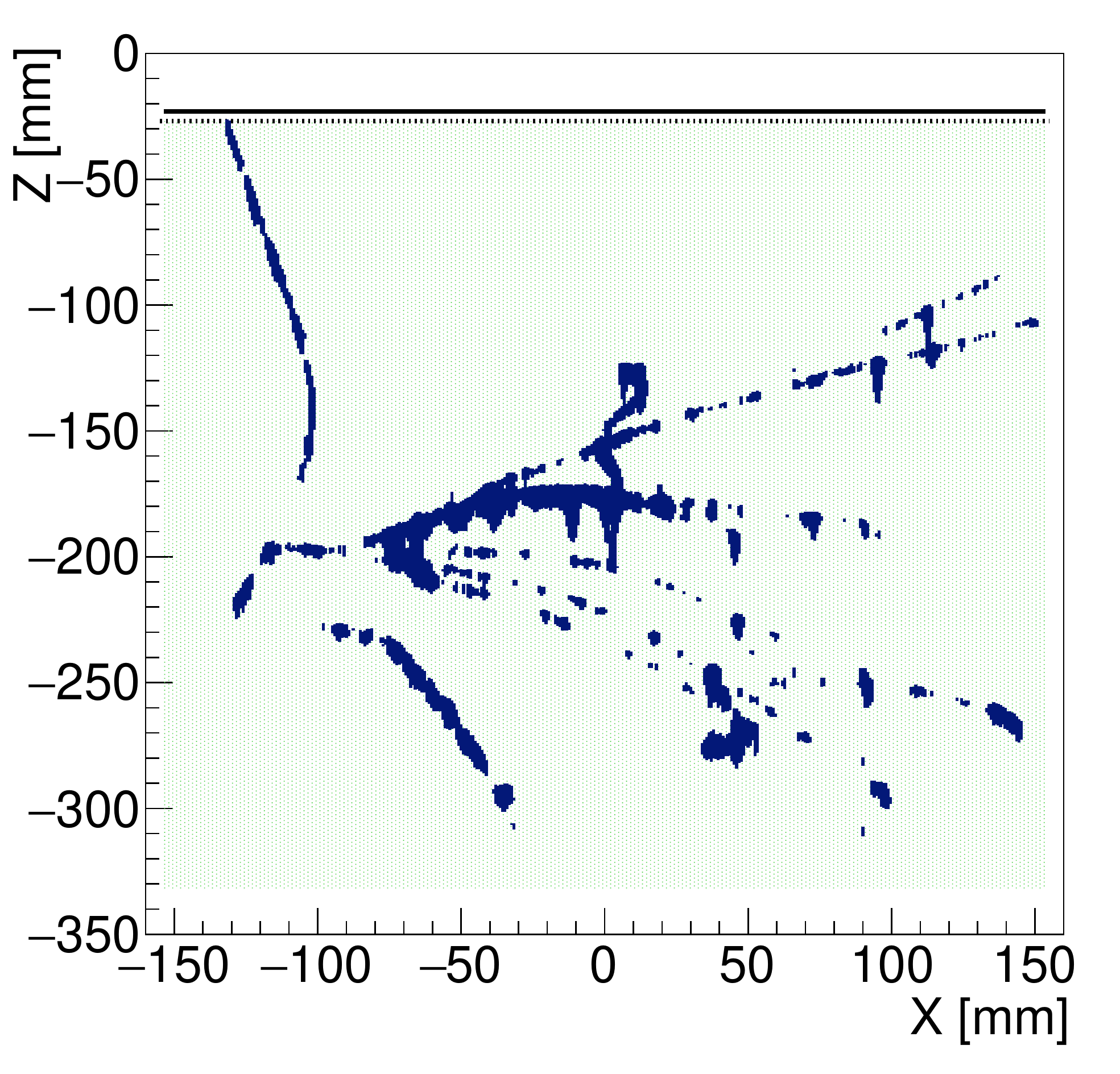}
    \caption{
        Typical tracks obtained by SMILE-2+ in the level flight.
        (top-left: single-electron event, top-right: pair-production event, 
        bottom-left: cosmic-ray event, bottom-right: shower event)
        The hatched area represents the active volume of the TPC,
        and the upper and lower sides of each image represent the zenith and nadir directions, respectively.
    }
    \label{fig:tracks}
\end{figure}

\begin{figure}
	\plotone{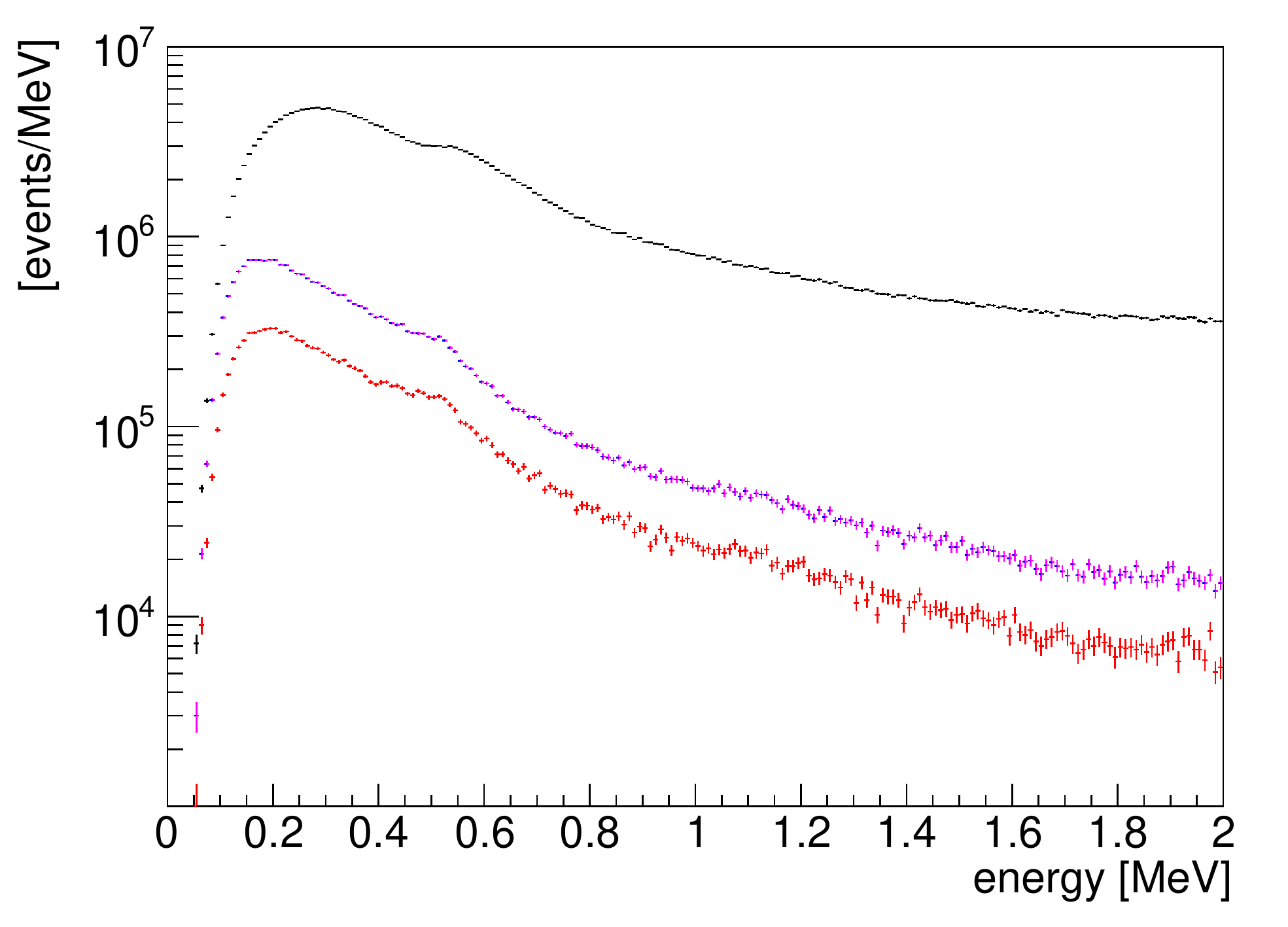}
	\caption{
		Total energy spectrum of each event selection during level flight 
		(see Section~\ref{sec:cal} for details).
		The black and magenta spectra were taken after selecting
		single pixel scintillator hits and fully-contained electrons, respectively,
		and the red spectrum was obtained by Compton scattering kinematics.
	}
	\label{fig:level_spec}
\end{figure}

SMILE-2+ ETCC recorded $4.9\times10^7$~events after turn-on. 
Typical tracks detected during level flight are shown in Fig.~\ref{fig:tracks}.
Single-electron events were selected for the gamma-ray reconstruction as described in Section~\ref{sec:cal}.
The measured track images clearly showed the occurrences in SMILE-2+ ETCC
from which select single-electron events were selected.
Thus, the track image itself provides a simple and powerful means of noise suppression.
At balloon altitudes, cosmic-rays interact with the structural materials to produce positrons or gamma-rays. 
Because the ETCC is triggered by coincident PSAs and TPC,
many cosmic-ray induced events including the electron-positron annihilation line absorbed in PSA were recorded. 
Using the cosmic-ray or shower events in the flight data, 
we corrected the gain of the scintillators by the annihilation line every 30~min 
and the TPC gain by the energy deposition rate of the minimum ionizing particles every 10~min.
Figure~\ref{fig:level_spec} shows the energy spectrum of each event selection during level flight,
and the spectrum obtained by selecting all events, when $2.4\times10^{5}$~events remained.
After selecting the fully-contained electrons, the spectrum showed a clear excess at 0.511~MeV. 

Figure~\ref{fig:air_mass} e) shows 
the light curve obtained by the gamma-ray reconstruction in all directions with the live-time correction.
The gamma-ray event rate in all directions was 2.7~events~s$^{-1}$ and remained stable during level flight.
When the air mass of the Galactic center is less than 4~g~cm$^{-2}$, 
the light curve rises with increasing elevation angle of the Galactic center.
The excess at the culmination time of the Galactic center is 0.5~events~s$^{-1}$, 
so the estimated intensity of the GCR is $\sim$0.15~photons~s$^{-1}$~cm$^{-2}$~sr$^{-1}$~MeV$^{-1}$
with an effective area of 1.1~cm$^2$ and FoV of 3.1~sr.
The intensity result is roughly consistent with the intensity measured SPI/{\it INTEGRAL}~\citep{2011ApJ...739...29B}.

\begin{figure}
    \plotone{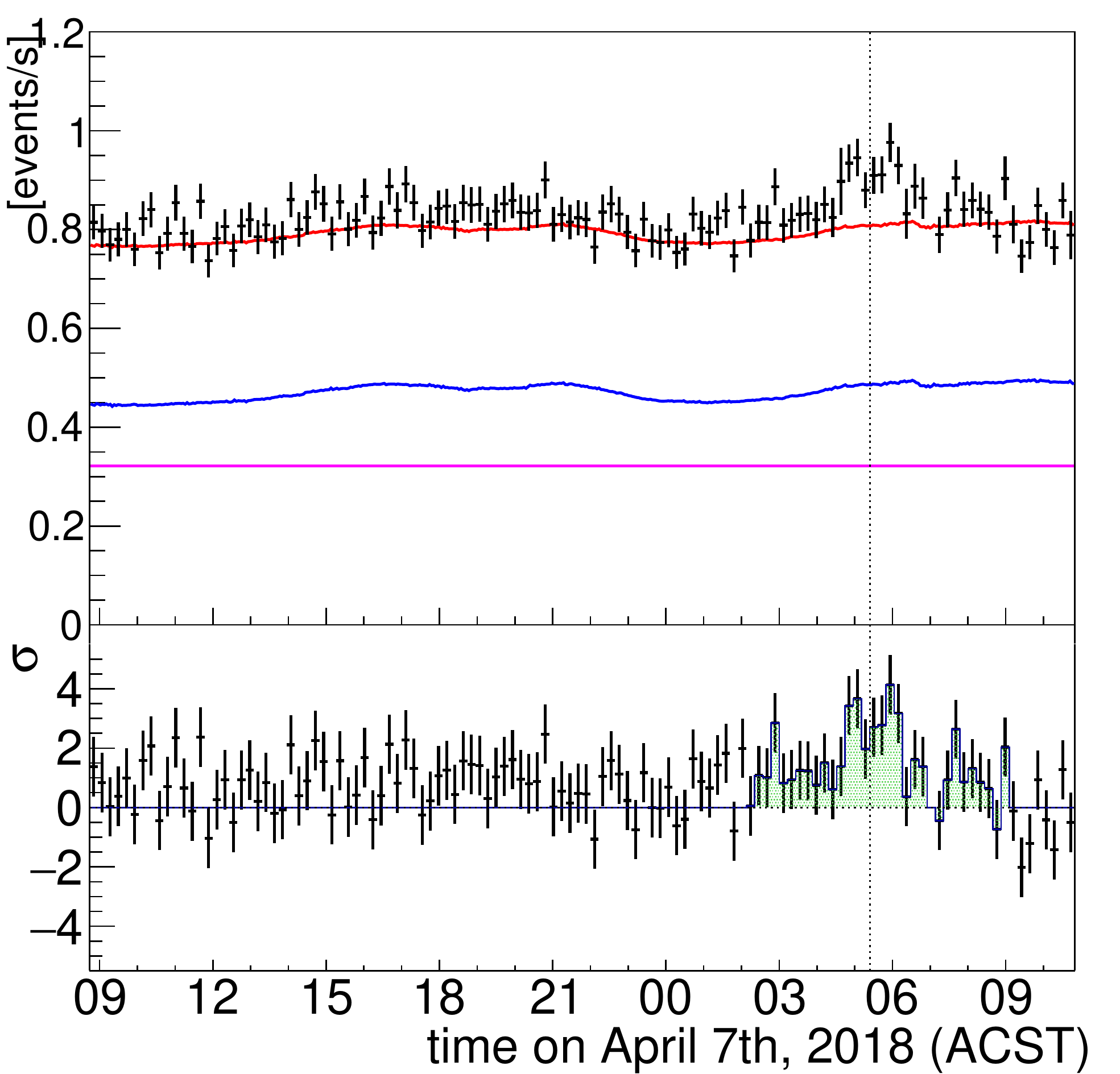}
    \caption{
        Light curve in the 0.15--2.1~MeV energy band at zenith angles below 60$^\circ$ during level flight.
        The vertical dotted line is the culmination time of the Galactic center.
        The blue curve describes the expected event rate 
        when SMILE-2+ detected extragalactic diffuse and atmospheric gamma-rays. 
        The magenta curve is the simulated background event rate induced 
        by protons, neutrons, electrons and positrons. 
        The red line describes the total estimated event rate of SMILE-2+.
        The lower panel shows the difference between the observed and expected event rates. 
        Within the hatched area, the air mass of Galactic center was less than 4~g~cm$^{-2}$.
    }
    \label{fig:rate_fov}
\end{figure}

\begin{figure}
	\plotone{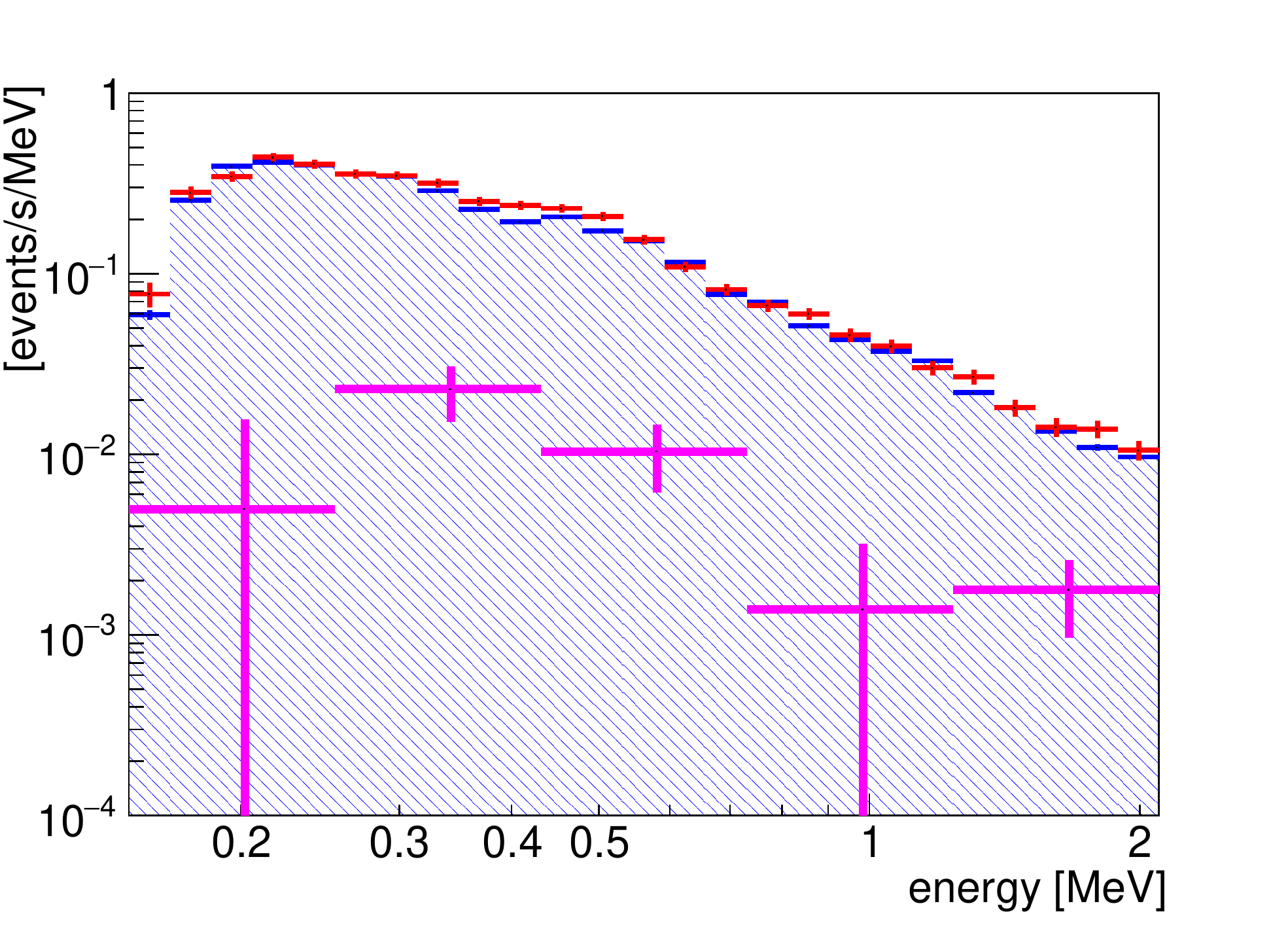}
	\caption{
		Observed energy spectrum of the ON-region (red), 
		estimated background (hatched area, see text), and the subtraction (magenta).
	}
	\label{fig:raw_spec}
\end{figure}

Figure~\ref{fig:rate_fov} shows the light curve at zenith angles below 60$^\circ$ 
in the 0.15--2.1 MeV energy band.
Plotted are the expected event rate at which SMILE-2+ detects the extragalactic diffuse and atmospheric gamma-rays (blue curve) and the expected instrumental background event rate 
induced by protons, neutrons, electrons and positrons (magenta line), 
which was calculated by PARMA~\citep{2008RadR..170..244S} and Geant4.
The intensity of atmospheric gamma-rays depends on the atmospheric depth, cutoff rigidity $R$, and solar modulation.
Here, we adopted the semiempirical model by \citet{1975JGR....80.3241L} 
to the atmospheric depth dependence of extragalactic diffuse and atmospheric gamma-rays. 
We additionally assumed that the intensity of atmospheric gamma-rays is proportional to $1.2 R^{-1.13}$, as detailed in Section~\ref{sec:discussion}.
The red line in Figure~\ref{fig:rate_fov} plots the total event rate of SMILE-2+
estimated independently of the observation results.
Over most of the level flight period, 
the observed rate 0.15--2.1 MeV events at a zenith angle of 60$^\circ$ is explained 
by the sum of extragalactic gamma-rays, atmospheric gamma-rays, and the instrumental background. 
The differential between the obtained and estimated event rates (lower panel of Figure~\ref{fig:rate_fov}) 
shows a significant excess around the culmination time of the Galactic center.
The detector count rates of SMILE-2+ clearly reveal an enhanced gamma-ray emission from the GCR.
In contrast, 
no clear excess appeared in Fig.~\ref{fig:air_mass} e) 
when the Crab nebula was observed.
This absence is explained by the low flux of the Crab nebula (only $\sim$3\% of the photon number of extragalactic diffuse gamma-rays in the FoV of the ETCC in the energy range 0.2--2.1~MeV).
During the ON-region of Crab nebula observation, 
we considered that the Crab nebula ($l$ = 184.6$^\circ$, $b$ = -5.8$^\circ$) was centered 
in a circle of radius 40~degrees, which defines the HPR of the PSF at 0.3~MeV. 
The spectrum obtained in the ON-region is the red spectrum in Fig.~\ref{fig:raw_spec}.

\subsection{background estimation}
\begin{figure}
    \plotone{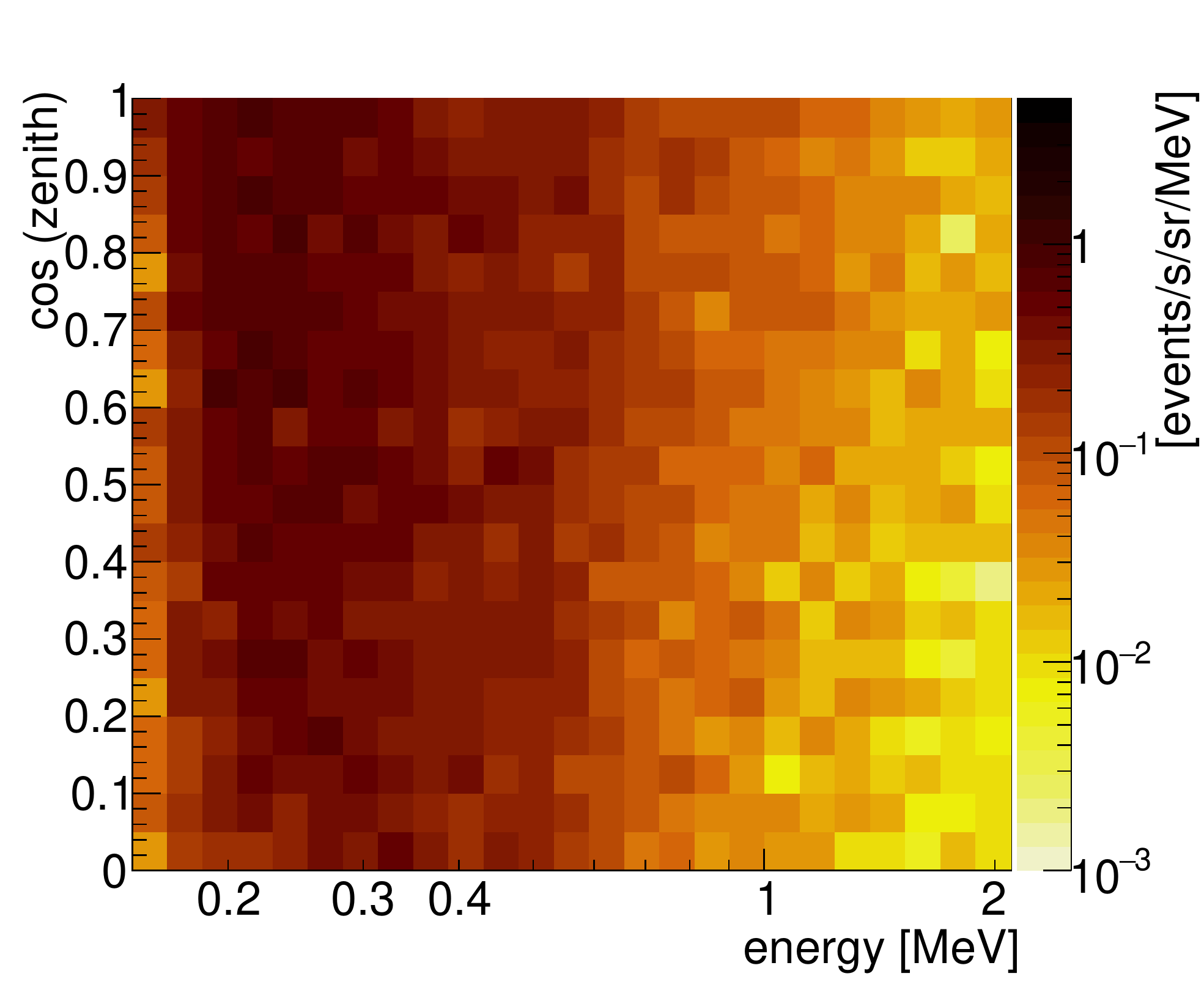}
    \caption{
	    Event intensity map $B(E', \theta')$ in the detected energy $E'$ 
	    versus zenith angle$\theta$ during the OFF-period. 
    }
	\label{fig:background}
\end{figure}

\begin{figure}
    \plotone{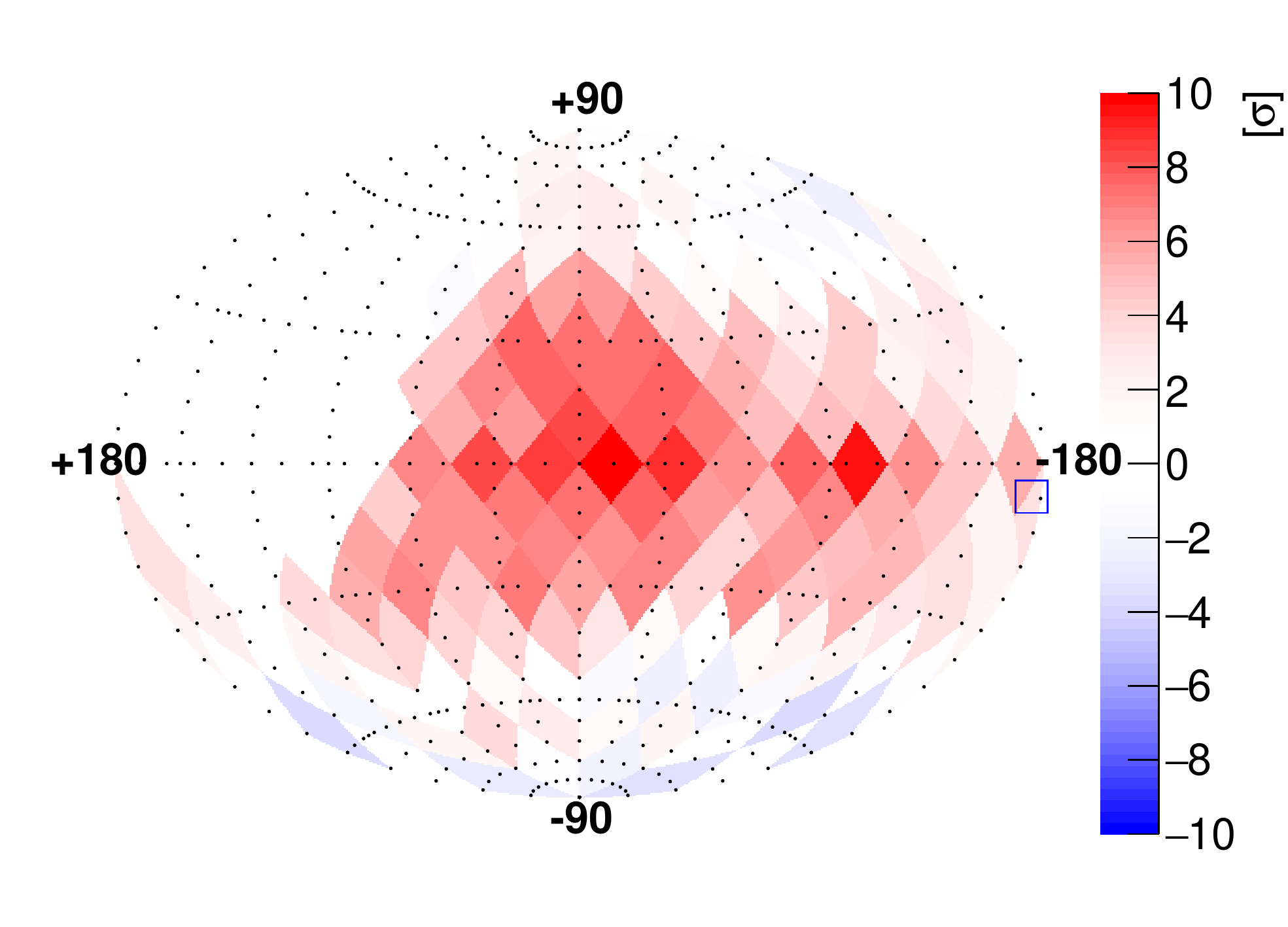}
    \caption{
        Significance survey map in galactic coordinates.
        The open square represents the Crab nebula.
    }
    \label{fig:sgm_map}
\end{figure}

In celestial gamma-ray observations at balloon altitudes, 
the gamma-ray background comprises extragalactic diffuse gamma-rays, 
atmospheric gamma-rays, and instrumental gamma-rays.
Although extragalactic diffuse gamma-rays are isotropic at the top of the atmosphere, 
they are scattered and attenuated in the atmosphere at balloon altitudes.
The intensities of atmospheric gamma-rays and the instrumental gamma-rays depend 
on the atmospheric depth, zenith angle, and intensity of cosmic-rays.
Unfortunately, the atmospheric depth increased by 20\% during observations of the Crab nebula. 
We thus defined an OFF-period of 20:30--22:50 ACST on April~7 of 2018,
when the balloon altitude decreased over the Crab observation period (Fig.~\ref{fig:altitude})
and no bright celestial objects appeared inside the FoV.
Figure~\ref{fig:background} maps the event intensity $B (E', \theta')$
on the detected energy $E'$ versus zenith angle $\theta'$ plane during the OFF period.
This event intensity map was assumed as the sky image of background gamma-rays in horizontal coordinates.
Because the zenith angle of the ON-region avaried over time, 
we calculated the average background energy spectrum $g (E')$ over the Crab observation period;
\begin{equation}
	g (E') = \frac{1}{T_{\text{obs}}} \int_{\text{ON-region}} B(E', \theta'(t)) d\Omega dt ,
	\label{eq:bg_rate}
\end{equation}
where $T_{\mathrm{obs}}$ is the live time of the Crab observation period
(hatched spectrum in Fig.~\ref{fig:raw_spec}).
After subtracting the estimated background $g (E')$ from the ON-region events, 
we obtained $f (E')$,
the energy spectrum of gamma-rays from the Crab nebula.
The magenta spectrum in Fig.~\ref{fig:raw_spec} displays $f (E')$ as a function of $E'$.
The convolved significance is 4.0$\sigma$.
Applying the same method to the observed data from 08:44 ACST on April 7 of 2018 to 06:30 ACST on the following day, 
we calculated the significance map shown in Fig.~\ref{fig:sgm_map}.
Because the balloon altitude varied in time, 
we defined $g(E')$ in two time period: from 10:30 to 13:30 ACST on April~7 of 2018,
when the atmospheric depth was less than 3.1~g~cm$^{-2}$, 
and from 20:30 to 22:50 ACST on April~7 of 2018, when the atmospheric depth exceeded 3.1~g~cm$^{-2}$.
In addition to the light curve,
SMILE-2+ significantly detected the GCR.
A detailed study of the GCR will be described elsewhere.

\subsection{calculation of flux}
\begin{figure}
    \plotone{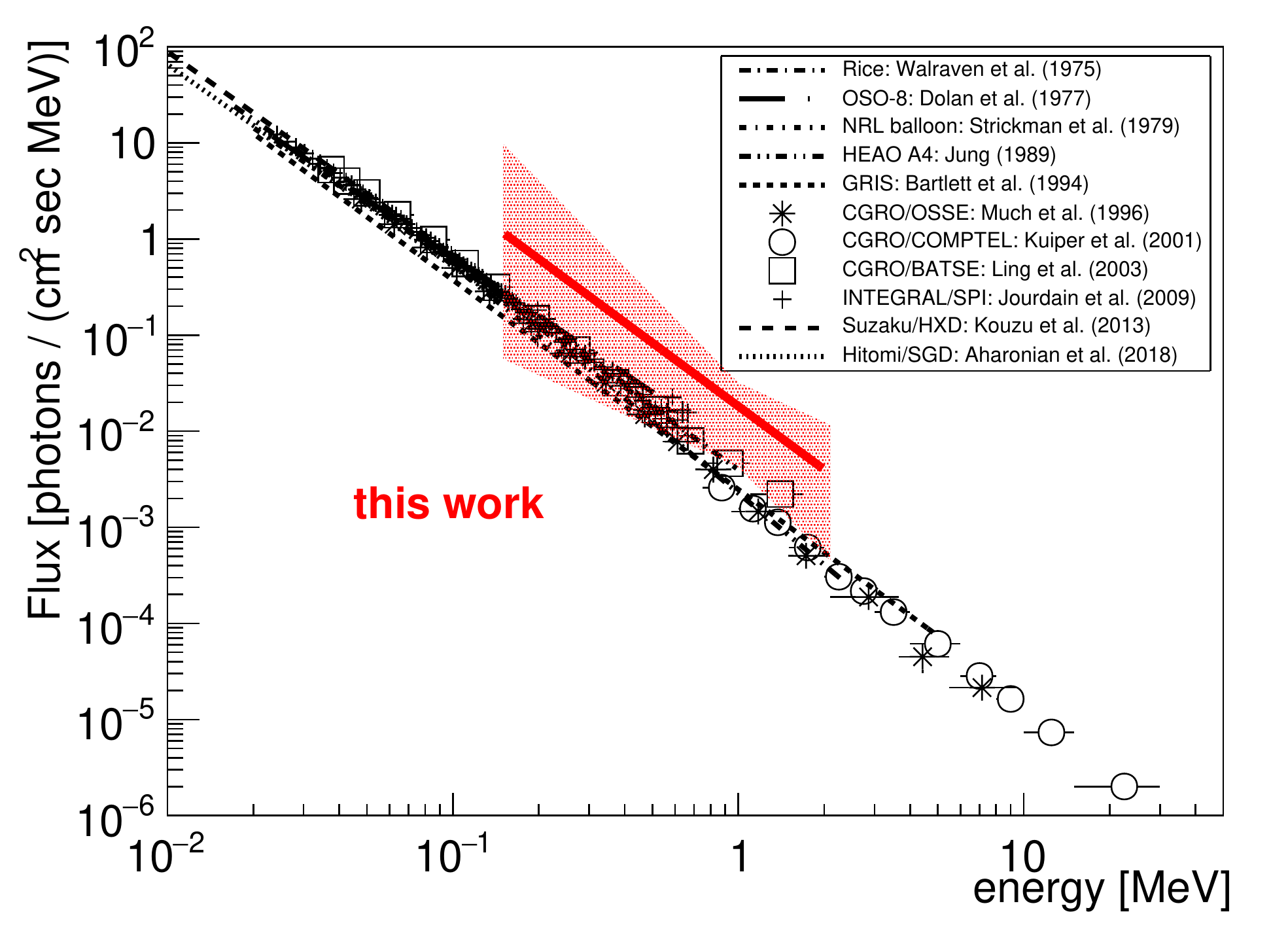}
    \caption{
	    Photon flux spectrum of the Crab nebula.
	    Solid line and hatched area are the best-fit 
	    and statistical error region (1$\sigma$), respectively, obtained by SMILE-2+.
    }
	\label{fig:crab_flux}
\end{figure}

The background-subtracted spectrum in Fig.~\ref{fig:raw_spec} includes the detector response and attenuation by the atmosphere,
\begin{equation}
	f (E') = \frac{1}{T_{\text{obs}}} \int f_c(E) A(E, \theta, E', \theta')
	\exp \left( -\frac{z \tau_{\text{tot}}}{\cos \theta } \right) dE dt ,
\end{equation}
where $E$ and $\theta$ are the true energy and true zenith angle of the incident photons, respectively,
$A$ {\color{blue}is the} response matrix of the ETCC,
and $z$, $\tau_{\mathrm{tot}}$, and $f_c$ denote
the atmospheric depth, cross section of total attenuation in the atmosphere,
and photon flux of the celestial object, respectively.
$A(E, \theta, E', \theta')$ can be estimated using the ETCC simulator described in Section~\ref{sec:cal}.
Because $\theta$, $\theta'$, and $z$ are temporally variable in balloon observations,
we calculated the time-averaged response matrix as follows:
\begin{equation}
	\bar{R}_{ij} = \frac{1}{T_{\text{obs}}} \int A(E_i, \theta, E'_j, \theta') \exp \left( -\frac{z \tau_{\text{tot}}}{\cos \theta} \right) dt ,
\end{equation}
where $i$ and $j$ are integers denoting specific energy bins.
The resultant spectrum is then described as
\begin{equation}
	f(E'_j) = \sum_i \bar{R}_{ij} f_c(E_i) ,
\end{equation}
where $f_c$ can be obtained thorough deconvolution.
Assuming that $f_c$ follows a single power-law, 
the deconvolved photon flux was detemined as 
$(1.82\pm1.40) \times 10^{-2} \left( E / \text{MeV} \right)^{-2.19 \pm 0.82}$ photons~s$^{-1}$~cm$^{-2}$~MeV$^{-1}$.
This flux spectrum is shown in Fig.~\ref{fig:crab_flux} together with its 1$\sigma$ error band.
The {\it Swift}/BAT transient monitor~\citep{2013ApJS..209...14K} revealed
no significant flares during the observation time\footnote{https://swift.gsfc.nasa.gov/results/transients/index.html}.
Therefore, as a consistency check, 
we can compare our result with other observations of the Crab nebula.
Our result is indeed consistent with previous observations of the Crab nebula.

To investigate any time-dependence of the OFF-period, 
we calculated the photon fluxes during two additional background periods.
10:30--13:30 ACST on April~7 of 2018 and the Crab observation period.
During the first of these period, the atmospheric depth was 3.01~g~cm$^{-2}$, 
13\% thinner than during the Crab observation period (14:30--19:30 ACST on April~7 of 2018). 
Therefore, the intensity of atmospheric gamma-rays should decrease and
the resulting gamma-ray flux of the Crab nebula based on this background period should be overestimated.
The selection of the Crab observation period is justified
because the Crab nebular's flux is negligibly smaller than 
the intensity of extragalactic diffuse and atmospheric gamma-rays within ETCC's FoV.
The single power-law spectra within these additional two background periods were calculated
as described for the original background period.
The obtained parameters are listed in Table~\ref{tab:param}.
As the parameters were quite simillar in each case, 
the obtained photon flux little depended on the selection of the OFF-period.
\begin{deluxetable}{cccc}
    \tablecaption{Photon flux parameters obtained in different background periods}
    \tablehead{
	\colhead{
	    BG time\tablenotemark{a}} & \colhead{20:30--22:50} & \colhead{10:30--13:30} & \colhead{14:30--19:30}
    }
    \startdata
    normalization\tablenotemark{b} & $1.82\pm1.40$ & $2.08\pm1.38$ & $1.55\pm1.37$ \\
    photon index  & $2.19\pm0.82$ & $2.28\pm0.81$ & $2.05\pm0.80$ \\
    significance & 4.0$\sigma$ & 6.6$\sigma$ & 2.9$\sigma$ \\
    \enddata
    \tablenotetext{a}{time on April 7, 2018 in ACST}
    \tablenotetext{b}{unit in $10^{-2}$ photons s$^{-1}$ cm$^{-2}$ MeV$^{-1}$}
    \label{tab:param}
\end{deluxetable}

\section{Discussion} \label{sec:discussion}
\begin{figure}
    \plotone{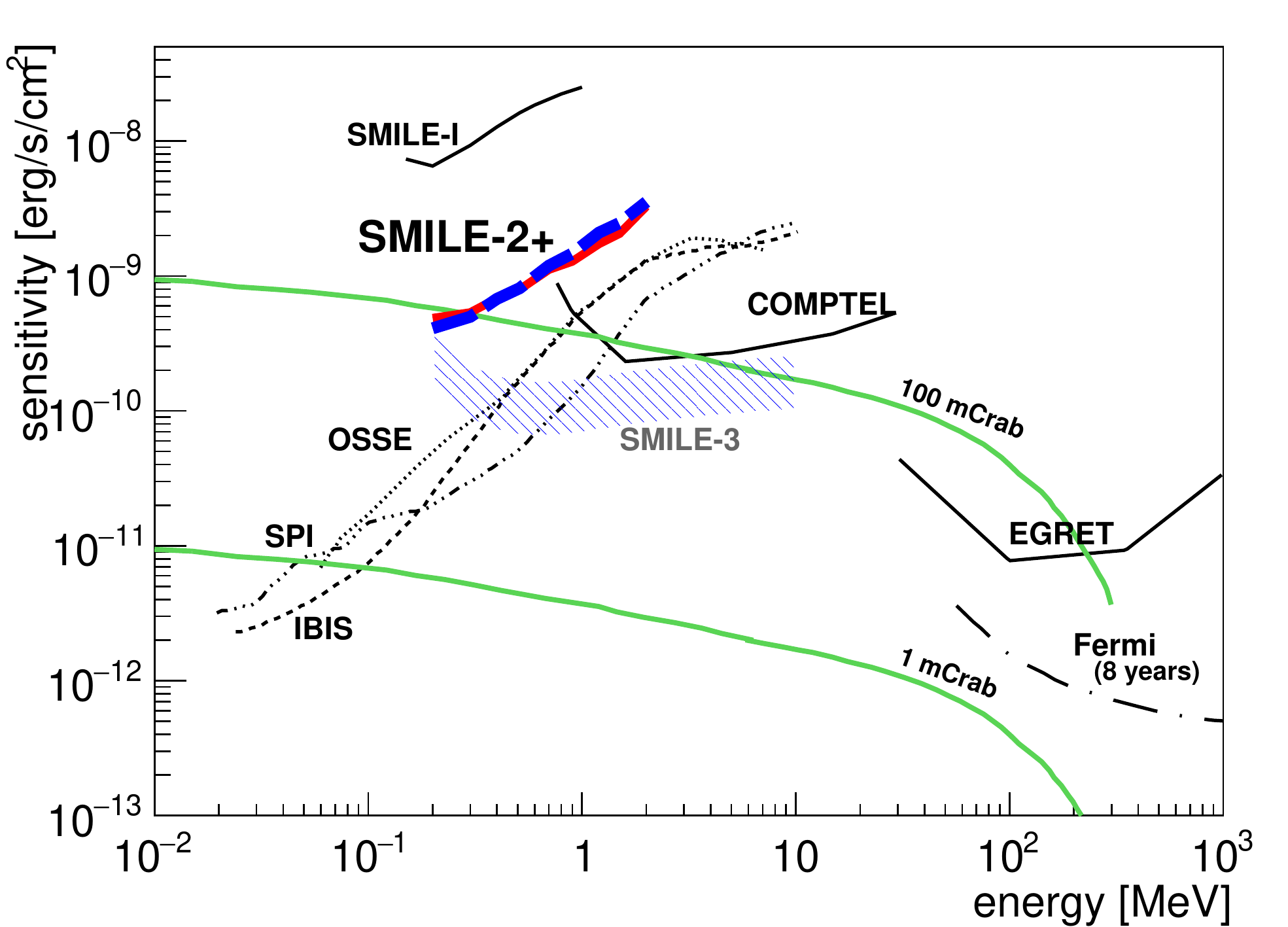}
    \caption{
	    3 $\sigma$ detection sensitivity of SMILE-2+ for the continuum spectrum 
	    at an observation time of $10^6$~s within an energy window of $\Delta E = E$.
	    The blue long-dashed line represents the realized sensitivity of SMILE-2+ based on the actual background event intensity.
	    The red solid line plots the sensitivity of SMILE-2+ against the background photons
	    comprising extragalactic diffuse gamma-rays, atmospheric gamma-rays, and the instrumental background.
	    The blue hatched area represents the estimated detection sensitivity
	    of next observation (SMILE-3). 
	    The uncertainty in the SMILE-3 sensitivity depends on the cutoff rigidity, balloon altitude, and solar modulation.
	    The black lines represent the results of previous studies 
	    \citep{2009ApJ...697.1071A, 2013APh....43..142T, 2011ApJ...733...13T}.
    }
	\label{fig:sens}
\end{figure}

In general, the detection sensitivity $S(E)$ in some energy band is defined 
by the detectable flux with a significance of 3$\sigma$, an observation time of 10$^6$~s, 
and an energy window of $\Delta E = E$.
The background rate and the detection sensitivity of SMILE-2+ are
respectively estimated by Eq.~(\ref{eq:bg_rate}) and by
\begin{equation}
	S(E) = \frac{3 \sqrt{T_{\text{obs}} \int B(E', \theta') dE' d\Omega' } }{T_{\text{obs}} \int A(E,0^\circ,E',\theta') dE' d\Omega'} ,
\end{equation}
in the zenith direction ($\theta$ = 0$^\circ$).
When $d\Omega$ is defined by the HPR (see in Fig.~\ref{fig:psf}),
the detection sensitivity of SMILE-2+ is the blue long-dashed line in Fig.~\ref{fig:sens}.
The realized sensitivity, at which the Crab nebula was detected within a few hours, 
was approximately ten times better than that of SMILE-I.

The background events comprise extragalactic diffuse gamma-rays, atmospheric gamma-rays, 
instrumental gamma-rays, and other particles (such as neutrons). 
Radioactivation is negligible because the duration time is too short in balloon observation.
Therefore the background event intensity $B(E', \theta')$ will be described as
\begin{equation}
	B(E', \theta') = \int (I_c + I_a) A(E, \theta, E', \theta') dE d\Omega + B_{\text{instr}},
\end{equation}
where $I_c$ and $I_a$ are intensities of extragalactic diffuse and atmospheric gamma-rays, respectively,
and $B_{\mathrm{instr}}(E', \theta')$ is the event intensity of the instrumental background.
The intensity of extragalactic diffuse gamma-rays depends on the zenith angle 
and changes with atmospheric attenuation and scattering 
at balloon altitudes \citep{1970Ap&SS...8..251M, 1971NPhS..229..148H, 1977ApJ...217..306S, 2011ApJ...733...13T}. 
In contrast, it is uniform at the top of the atmosphere.
Atmospheric gamma-rays, produced when cosmic-rays interact with the atmosphere, 
depend on the atmospheric depth, cutoff rigidity, and solar modulation.
Among several models for atmospheric gamma-rays is PARMA \citep{2008RadR..170..244S},
an analytical model based on PHITS simulations \citep{PHITS_2002}.
PARMA models the intensity of atmospheric gamma-rays
with respect to energy, zenith angle, atmospheric depth, solar modulation, and cutoff rigidity,
but does not consider the primary cosmic electrons/positrons as the initial particles.
\citet{1975JGR....80.3241L} and \citet{1977JGR....82.1463L} included extragalactic diffuse gamma-ray
in a semiempirical model based on balloon observations.
Their model considers the gamma-ray energy, zenith angle and atmospheric depth,
but not consider the effect of cutoff rigidity and solar modulation on gamma-ray intensity.
Nevertheless, this model is often cited in observational studies.

\begin{figure}
    \plotone{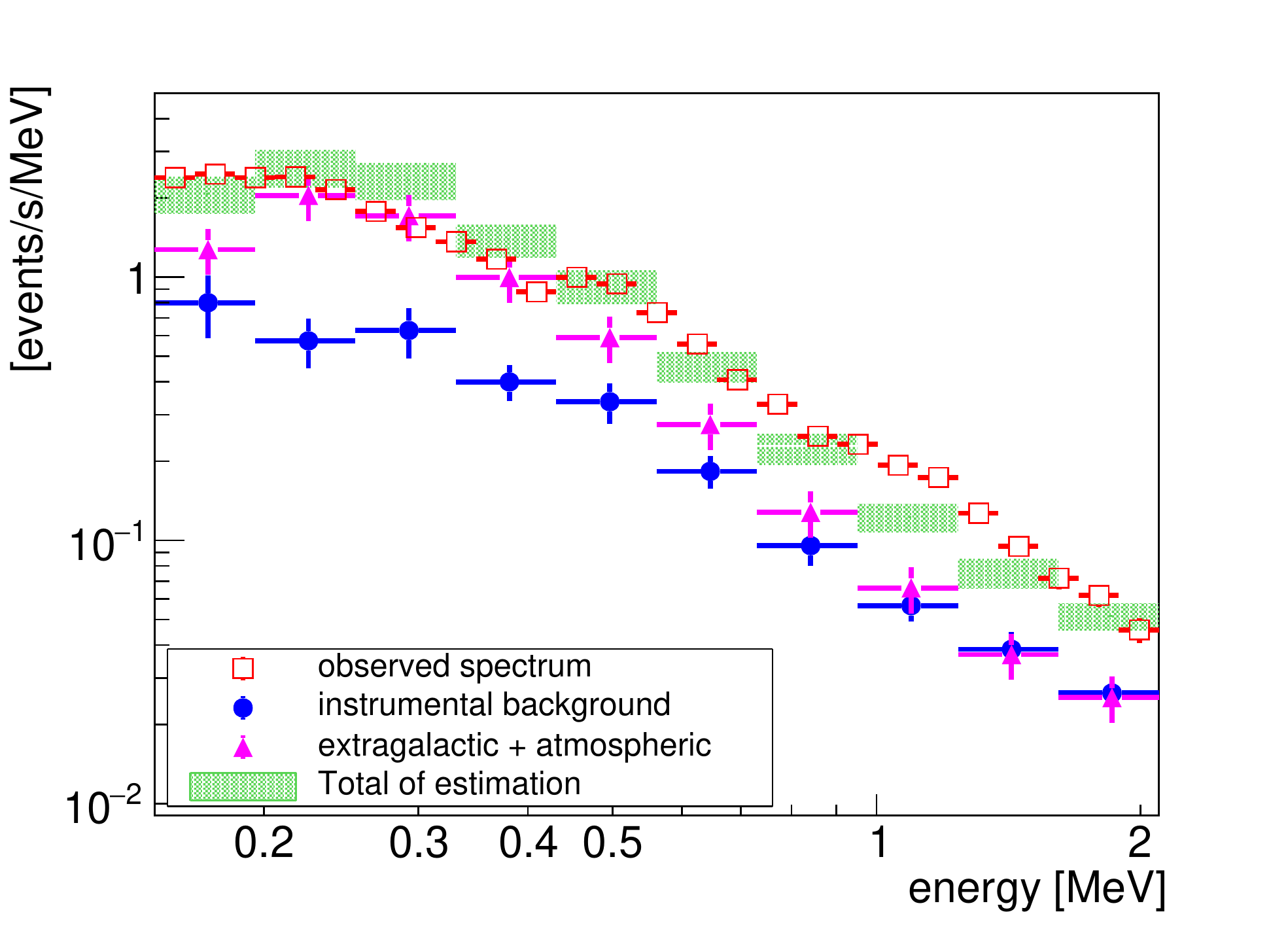}
    \caption{
        Energy spectra obtained within the zenith angle of 60~degrees at an atmospheric depth of 3.0~g~cm$^{-2}$.
        Open squares plot the events detected during 10:30--13:30 ACST on April~7 (ACST) of 2018.
        The filled circles, filled triangles and hatched areas represent the estimated instrumental background, 
        summation of the extragalactic diffuse and atmospheric gamma-ray components based on the Ling model
        \citep{1975JGR....80.3241L, 1977JGR....82.1463L}, 
        and the total estimated spectrum, respectively.
    }
    \label{fig:bg_spec}
\end{figure}

As the intensity model of extragalactic diffuse and atmospheric gamma-rays, 
we adopted the \citet{1975JGR....80.3241L} and \citet{1977JGR....82.1463L} model 
and scaled by the dependences of cutoff rigidity and solar modulation.
To correct the cutoff rigidity, we scaled the atmospheric component in the models by $(4.5 \ \text{GV}/8.4 \ \text{GV})^{1.13}$.
Meanwhile, the balloons referred by Ling's models were launched around the solar maximum, 
whereas SMILE-2+ was launched near the solar minimum.
As the atmospheric gamma-ray intensity is around 1.2--2.0 times 
larger at the solar minimum than at the solar maximum \citep{1984JGR....8910685M, 2007MNRAS.377.1726S},
we scaled the intensity of atmospheric gamma-rays by a factor of1.2.
Assuming the gamma-ray intensity at an atmospheric depth of 3.0~g~cm$^{-2}$, 
we expected the SMILE-2+ ETCC spectrum represented by the filled triangles in Fig.~\ref{fig:bg_spec}.
To evaluate the instrumental background $B_{\mathrm{intsr}}(E', \theta')$, 
we simulated the reconstructed events of the ETCC simulator
using the PARMA-calculated intensities of the initial particles (protons, neutrons, electrons, and positrons).
The filled circles in Fig.~\ref{fig:bg_spec} describe the estimated energy spectrum of the instrumental background at the same altitude.
The amount of instrumental background was one-half that of the essential background composed of extragalactic diffuse and atmospheric gamma-rays.
Thus the instrumental background negligibly affected the detection sensitivity of SMILE-2+.
The expected total background spectrum (indicated by the hatched area in Fig.~\ref{fig:bg_spec})
approximately matched with the observed energy spectrum (described by the open squares in Fig.~\ref{fig:bg_spec}).
Against this estimated total background, 
we can obtain the detection sensitivity independently of the sensitivities mentioned above.
The estimated detection sensitivity of SMILE-2+ (red solid line in Fig.~\ref{fig:sens}) 
was also consistent with the realized sensitivity.

As confirmed in these results,
the background spectrum and the detection sensitivity of observations by an ETCC in space
are well embodied in the results of the ground calibrations.
In contrast, 
the obtained sensitivities of COMPTEL and most of the other conventional Compton cameras are 
several times worse than the expected sensitivities \citep{2004NewAR..48..193S, 2011ApJ...738....8B}.
The above estimationis strongly and independently supported 
by the enhanced light curve at the culmination time of the Galactic center.
If the instrumental background was several times more intense than the extragalactic gamma-rays,
such as COMPTEL,
this enhancement could not be observed.
The instrumental background of ETCC observations is
mainly contributed by gamma-rays generated by interactions between cosmic rays and the instrumental material.
Other disturbances are diminished or removed by powerful background-rejection tools 
such as particle identification, track image, and a Compton kinematics test.
The upward instrumental gamma-rays are intense because of the instrumental material is below the ETCC.
However, because the ETCC can pinpoint the direction of an incident gamma-ray to a point in each event. 
SMILE-2+ can suppress the instrumental background within the FoV to the expected contamination level the PSF.
Under this noise condition, the detection sensitivity of the next observation (SMILE-3)
can be estimated both easily and realiably.
After constructing a new TPC of volume $30 \times 30 \times 30$~cm$^3$ 
filled with CF$_4$ gas at a 3~atm pressure and optimizing the structural design, 
the effective area of ETCC will be approximately 10~cm$^2$ for 0.4~MeV and 1~cm$^2$ for 2.5~MeV.
To improve the PSF, we are developing an analysis method based on machine learning. 
In the first trial, deep learning of the recoil direction and scattering point improved  
the PSF to double that of the present paper \citep{2021arXiv210502512I}.
If the direction accuracy of Compton-recoil electrons could be improved to the limiting accuracy of multiple scattering, 
the PSF of the ETCC will improve to 10 degrees and 2~degrees at HPR for 0.4~MeV and 2.5~MeV, respectively.
The sensitivity of an ETCC loaded on a long-duration balloon (SMILE-3)should exceed
that of COMPTEL by several times (indicated by the blue hatched area in Fig.~\ref{fig:sens}).
Recently, a super-pressure balloon has been launched 
for over one month at middle latitudes in the southern hemisphere \citep{2016int..workE..75K}.
{color{blue}Therefore, if we can launch an updated ETCC on a long-duration balloon in the SMILE-3 mission, 
we could significantly surpass the COMPTEL observations
because the large FoV of the ETCC enables observation times of $\sim$10$^6$~s. }

\section{Summary}
To advance MeV gamma-ray astronomy, 
we are developing an ETCC with imaging spectroscopy ability. 
The ETCC is based on bijection imaging and utilizes powerful background rejection tools 
(particle identification, Compton-scattering kinematics test, and charged-particle track imaging).
A proper PSF is performed on the celestial sphere,
enabling acquisition of the energy spectrum of the observation target by simple ON-OFF method.
In 2018, we launched the second balloon (SMILE-2+) to confirm ETCC observations of celestial objects.
The effective area of the ETCC loaded on SMILE-2+ is 1.1~cm$^2$ for 0.356~MeV,
the PSF is 30~degrees at HPR for 0.662~MeV, and the FoV is 3.1~sr
when a 30$\times$30$\times$30~cm$^3$ TPC is filled with argon gas at 2~atm pressure.
The observed flux and energy spectrum during the level fight,
estimated independently of the experimental results, 
were well explained by a background of extragalactic diffuse, atmospheric, and instrumental gamma-rays.
With this good understanding of the sensitivity and background,
we achieved the pre-launch expectations (a significance level of 4.0$\sigma$) of SMILE-2+ observations
of gamma-rays from the Crab nebula.
The obtained flux was also consistent with other flux observations.
In addition, the light curve and the significance survey map show 
that the GCR is very bright 
with a significance of $\sim$10$\sigma$ in the 0.2--2.1 MeV energy range.
Thus, SMILE-2+ is the first application of imaging spectroscopy 
based on a proper PSF and bijection imaging to MeV gamma-ray astronomy.
The observed energy spectrum during level flight was explained 
by the background noise, which contains extragalactic diffuse, atmospheric, and instrumental gamma-rays.
The achieved detection sensitivity of SMILE-2+ matched the sensitivity estimated from ground calibrations,
whereas the sensitivities of most conventional Compton cameras are several times worse than their expected values.
The instrumental gamma-rays little affected the detection sensitivity of SMILE-2+, 
because they constituted only one-third of the background.
For this reason, the ETCC overcomes the large background problem. 
When designing a Compton camera, the sensitivity must be estimated with a PSF (not an ARM),
similarly to telescopes operating in the X-ray or GeV bands.
In the near future, 
the ETCC will be updated to an approximate effective area of 10~cm$^2$ and a PSF of several degrees at the HPR,
and we will then launch a long-duration balloon flight (SMILE-3) for scientific observations.
The ETCC can become a unique pioneer with deeper survey ability than COMPTEL in MeV gamma-ray astronomy.

\acknowledgments
The balloon-borne experiment was conducted by Scientific Ballooning (DAIKIKYU) Research and Operation Group, ISAS, JAXA.
This study was supported by the Japan Society for the Promotion of Science (JSPS) 
Grant-in-Aid for Scientific Research (S) (21224005), (A) (20244026, 16H02185), 
Grant-in-Aid for Young Scientists(B) (15K17608), 
JSPS Grant-in-Aid for Challenging Exploratory Research (23654067, 25610042, 16K13785, 20K20428), 
a Grant-in-Aid from the Global COE program “Next Generation Physics, 
Spun from Universality and Emergence” from the Ministry of Education, Culture, Sports, Science and Technology (MEXT) of Japan, 
and Grant-in-Aid for JSPS Fellows (16J08498, 18J20107, 19J11323). 
Some of the electronics development was supported by KEK-DTP and Open-It Consortium.
And we thank Enago (www.enago.jp) for the English language review.

\end{document}